\documentclass[11pt]{amsart}
\usepackage[a4paper,margin=23mm,top=30mm]{geometry}

\usepackage{amsfonts, amsmath, amssymb, amsgen, amsthm, amscd}
\usepackage{newtxtext,newtxmath}
\usepackage[utf8]{inputenc} 

\usepackage{color}

\usepackage[all]{xy}
\usepackage{tikz-cd}
\usepackage{hyperref}
\usepackage{mathtools,slashed}
\usepackage{setspace}
\usepackage{enumerate}

\def\a{\alpha}

\def\d{\delta}

\def\g{\gamma}

\def\Om{\Omega}

\def\t{\theta}

\def\p{\partial}

\def\Om{\Omega}

\def\Ad{\mathop{\rm Ad}\nolimits}
\def\ad{\mathop{\rm ad}\nolimits}

\def\dt{\left.\frac{d}{dt}\right|_{_{t=0}}}

\usepackage{enumerate}

\newcommand{\G}[1]{\mathfrak{#1}}

\newcommand{\B}[1]{\mathbb{#1}}

\numberwithin{equation}{section}

\newtheorem{theorem}{Theorem}[section]

\theoremstyle{definition}

\newtheorem{remark}[theorem]{Remark}

\usepackage{tikz-cd}
\usepackage{adjustbox}
\usepackage{graphicx}
\usepackage[toc,page]{appendix}
\usepackage{color}

\begin{document}
	
	\title{Tulczyjew's Triplet with an Ehresmann connection I: Trivialization and Reduction}
	\author{O\u{g}ul Esen}
	\thanks{Department of Mathematics, Gebze Technical University,  41400 Gebze-Kocaeli, Turkey}
	\thanks{oesen@gtu.edu.tr}
	
	\author{Mahmut Kudeyt}
	\thanks{Department of Mechatronics Engineering, Kadir Has University,  34083 Cibali-\.{I}stanbul, Turkey}
	\thanks{mahmut.kudeyt@khas.edu.tr}
	
	\author{Serkan Sütlü}
	\thanks{Department of Mathematics, I\c{s}{\i}k University, 34980 \c{S}ile-\.{I}stanbul, Turkey}
	\thanks{serkan.sutlu@isikun.edu.tr}
	\date{}

	\begin{abstract}
We study the trivialization and the reduction of the Tulczyjew's triplet, in the presence of a symmetry and an Ehresmann connection associated to it. We thus obtain trivializations and reductions of iterated tangent and cotangent bundles $T^*TQ$, $TT^*Q$ and $T^*T^*Q$. Accordingly, the symplectomorphisms between these manifolds are properly trivialized and reduced. 
	\end{abstract}
	\maketitle
	
	\tableofcontents
	\onehalfspacing
	
	\setlength{\parskip}{7mm}
	
	\section{Introduction}
	
	Euler-Lagrange equations governing the motion of a physical system, whose configuration space is a finite dimensional manifold,  is determined by a Lagrangian function defined on the tangent bundle. On the other hand, Hamilton's equation is generated by a Hamiltonian function which is defined on the cotangent bundle \cite{abraham1978foundations,arnold1989mathematical,de2011methods, goldstein1980classical,libermann2012symplectic}. 
	In order to find the (Legendre) transformation  between the Lagrangian and the Hamiltonian realizations of a system, it is critical to know whether the Lagrangian / Hamiltonian function is (hyper)regular. In the case of a regular Lagrangian / Hamiltonian, the Legendre transformation is immediate, being the fiber derivative of the generating function. However, if the Lagrangian / Hamiltonian is singular, then establishing a relationship between the two formalisms is far from straightforward. 
	
To this end, a geometric framework (called the Tulczyjew's triplet) was proposed by Tulczyjew, which allows the Legendre transformation even for the singular and/or constrained systems,  \cite{tulczyjew1972hamiltonian,tulczyjew1976soush,tulczyjew1976sous,Tu77,Tu80,tulczyjew1989geometric,tulczyjew2004homogenous,tulczyjew1999slow}. The Tulczyjew's triplet is a construction that involves the (second order) iterated bundles of the configuration space, say $Q$, connected by two symplectomorphisms $\a_Q:TT^*Q\to T^*TQ$ and $\Om_Q^\flat:TT^*Q\to T^*T^*Q$, as well as their projections onto the (first order) (co)tangent bundles. In short, a Tulczyjew's triplet may be pictured by the diagram
\begin{equation} \label{TT-}
\xymatrix{T^{\ast }TQ \ar[dr]_{\tau^*_{TQ}}&&TT^{\ast}Q\ar[dl]^{T\tau^*_{Q}}
\ar[rr]^{\Omega_{Q}^{\flat}} \ar[dr]_{\tau_{T^{\ast}Q}} \ar[ll]_{\alpha_{Q}}&&T^{\ast }T^{\ast}Q\ar[dl]^{\tau^* _{T^{\ast}Q}}\\
&TQ & &T^{\ast}Q}
\end{equation}
In this theory, the dynamics of the system under consideration (whether it is in Lagrangian or Hamiltonian form) is formulated as a Lagrangian submanifold of the Tulczyjew symplectic space $TT^{\ast}Q$, that is, the tangent bundle of the momentum phase space. This is achieved by two special symplectic structures which are referred as the left wing and the right wing of the triplet, \cite{Be11,SnTu72,tulczyjew2004homogenous}. The Legendre transformation then is said to be established, if the Lagrangian submanifolds generated by a Hamiltonian function and a Lagrangian function coincide. For singular systems, one may need to refer to a Morse family in order to achieve the coincidence of the Lagrangian submanifolds.  
	
	
	The Tulczyjew's triplet has been constructed for many physical systems and on several different geometric theories. For instance, we refer the reader to \cite{EsGuSu20,esen2014tulczyjew,esen2015tulczyjew,grabowska2016tulczyjew} for the triplet over a Lie group (from the point of view of which the present paper is a generalization). In the case of the higher order dynamics, we may cite \cite{deLeLa89}, while for the field theories we refer the reader to \cite{de2003tulczyjew,echeverria2000geometry,grabowska2012tulczyjew, grabowska2013tulczyjew,roman2007k}. As for the higher order field theories, one may consult \cite{grabowska2015tulczyjew}. On the other hand, we refer the reader to \cite{bruce2015higher,grabowski2015new,GrGrUr14,grabowski1997tangent} for the graded bundles, and to \cite{jozwikowski2017prolongations} for a discussion related to prolongations. The extension of the Tulczyjew’s triplet to the level of Lie algebroids has been proposed in \cite{IgMaPaSo06}, see also \cite{abrunheiro2018lagrangian,bruce2010tulczyjew,deLeMaMa05,grabowska2006geometrical}. 
	
As one may find its roots in \cite{poincare1901forme}, the Lagrangian reduction theory \cite{bloch1996nonholonomic,n2001lagrangian,marsden1993lagrangian,scheurle1993reduced} is rather more recent compared to the Hamiltonian reduction theory \cite{marsden1974reduction,MaMiOrPeRa07,marsden1986reduction,marsden2001comments,meyer1973symmetries}. In the Lagrangian framework, the reduced dynamics on the orbit space of the tangent bundle of the configuration space is governed by the Lagrange–Poincaré equations. In the case of the configuration space to be the symmetry group, the Lagrange–Poincaré equations reduce further to the Euler-Poincaré equations on the Lie algebra of the symmetry group. As for the Hamiltonian perspective, the reduced dynamics is determined by the Hamilton-Poincaré equations on the orbit space of the cotangent bundle, which reduce to the Lie-Poisson equations if in particular the configuration space is the symmetry group.	

In the present note we intend to address both the trivialization and the reduction of the classical Tulczyjew's triplet in view of an Ehresmann connection. To be more precise, let us first recall that the symmetry of a physical system, in its very classical sense, may be given by a Lie group action on the configuration space of the system, \cite{holm2008geometric,holm2009geometric,MaMoRa90,MarsdenRatiu-book,Olv-book,olver1995equivalence}.  As such, we let the configuration space $Q$ be equipped with a free and proper action of a Lie group $G$. This way, we may regard $Q$ as a principal $G$-bundle over the orbit space $\bar{Q}:=G\backslash Q$, and hence we may endow the tangent bundle $TQ$ with a principal (Ehresmann) connection $A:TQ\to \G{g}$. The thus obtained decompositions  
\begin{equation} \label{decomp-intro}
TQ\simeq Q\times_{\bar{Q}} T\bar{Q}\times \mathfrak{g}, \qquad T^*Q\simeq Q\times_{\bar{Q}} T^*\bar{Q}\times \mathfrak{g}^*,
\end{equation}   
into Whitney sums (of the horizontal and the vertical subbundles), see for instance \cite{cendra2003variational,n2001lagrangian,MaMiOrPeRa07}, allow at once to express the $G$-reductions of both the tangent and the cotangent bundles as   
\begin{equation} \label{reduc-intro}
G\backslash TQ\simeq T\bar{Q}\times_{\bar{Q}} \tilde{\mathfrak{g}}, \qquad G\backslash T^*Q\simeq T^*\bar{Q}\times_{\bar{Q}} \tilde{\mathfrak{g}}^*,
\end{equation}
where $\tilde{\mathfrak{g}}$ denotes the adjoint bundle, while $\tilde{\mathfrak{g}}^*$ stands for the coadjoint bundle. These identifications are important not only for the geometrical considerations, but also for the concrete physical applications; they provide a decomposition of the dynamics into the vertical and the horizontal components. What is further, as we present hereby, the decompositions in \eqref{decomp-intro} and \eqref{reduc-intro} may be lifted to all second order iterated bundles $TTQ$, $T^*TQ$, $TT^*Q$ and $T^*T^*Q$, allowing to obtain the trivializations of the symplectomorphisms $\a_Q:TT^*Q\to T^*TQ$ and $\Om_Q^\flat:TT^*Q\to T^*T^*Q$, as well as all the projections on \eqref{TT-}.  

Recent works addressing the reduction of the Tulczyjew's triplet under symmetry include \cite{barbero2016inverse,garcia2014reduced,garcia2014geometric}, yet, none concerns the trivialization by virtue of a connection form. From this point of view, the present approach, along the lines of \cite{cendra2003variational,n2001lagrangian,MaMiOrPeRa07},  
allows to analyze the dynamics explicitly over the orbit spaces.  
		
\subsubsection*{The outline of the paper}	  

The paper is planned over three main sections. In order to establish the notation, conventions, and the terminology, we begin with the preliminaries in Section \ref{sect-prelim}. In particular, the very basic terminology regarding symplectic structures on cotangent bundles is given in Subsection \ref{subsect-symp-str}, while an overview of Tulczyjew triplets may be found in Subsection \ref{subsect-Tulczyjew-triplet}. What is more in preliminaries is a short discussion on the theory of (principal Ehresmann) connections in Subsection \ref{subsect-connection-curvature}. We then conclude this section with Subsection \ref{subsect-triv-reduct-tangent-cotangent} on the presentations of both the trivializations and reductions of (co)tangent bundles in view of an Ehresmann connection. The latter discussion is then lifted to the level of second order iterated (co)tangent bundles in Section \ref{Sec-TTQ-TT*Q}. To be more precise, the trivializations and the reductions of the iterated tangent bundles $TTQ$ and $TT^*Q$ are explicitly presented in Subsection \ref{Sec-tri-TTQ} and Subsection \ref{Sec-tri-TT*Q} respectively, while the iterated cotangent bundles $T^*TQ$ and $T^*T^*Q$ are treated in Subsection \ref{Sec-tri-T*TQ} and Subsection \ref{Sec-tri-T*T*Q}. The trivializations and the reductions of the rest of the components of the Tulczyjew's triplet, namely the symplectomorphisms, are postponed in Section \ref{Red-Trv-ssp}. More precisely, in Subsection \ref{subsect-canonical-forms} we consider the canonical 1-form and the symplectic 2-form on the cotangent bundle, while we undertake the task to trivialize (and then to reduce by the symmetry) the symplectomorphisms in Subsection \ref{subsect-symplecto}. On the other hand, Subsection \ref{subsect-symplect-sp} is reserved for the trivialization and the reduction of the Tulczyjew's symplectic space $TT^*Q$. 
	
\subsubsection*{Notations and Conventions} Throughout the text $Q$ will denote a smooth manifold admitting a (free and proper) left action $\phi:G\times Q\to Q$, $(g,q)\mapsto g\cdot q = \phi_g(q)$, of a Lie group $G$. We shall adopt the notation $\bar{Q}:=G\backslash Q$ for the orbit space, and $(Q,\pi,\bar{Q})$ for the principal $G$-bundle $\pi:Q\to \bar{Q}$. We shall also use the notations $\tau_M:TM\to M$ and $\tau^*_M:T^*M\to M$ to denote the tangent and cotangent bundle projections of a manifold $M$, respectively. On the other hand, given a manifold $M$, we shall use $\Lambda^k(M)$ to represent the space of $k$-forms on $M$, and $\G{X}(M)$ to denote the space of vector fields on $M$.

\section{Preliminaries}\label{sect-prelim}

In order to fix the terminology that we shall use in the sequel, we find it instructive to devote a section to brief surveys on the basic concepts to be used in the text. 

\subsection{Symplectic structures} \label{subsect-symp-str}~

Let us recall, very briefly, the canonical symplectic structure on a cotangent bundle. We shall then review the (special) symplectic structures on symplectic manifolds.

\subsubsection*{\rm \bf Cotangent bundles}

Given a manifold $Q$, the \emph{tautological (Liouville, or canonical) 1-form} $\theta_Q\in \Lambda^1(T^*Q)$ is given, on a vector field $X\in \G{X}(T^{*}Q)$, by
\begin{equation} \label{can-Lio}
\theta_Q(X)= \langle \tau_{T^{*}Q}(X), T\tau^*_Q(X)\rangle,
\end{equation}
where $\tau_{T^{*}Q}:TT^{*}Q \to T^{*}Q$ is the tangent bundle projection, and $T\tau^*_Q:TT^{*}Q \to TQ$ is the tangent mapping of the cotangent bundle projection. The negative of the exterior derivative of the canonical 1-form, that is $\Omega_Q:=-d\theta_Q$, then, represents the canonical symplectic 2-form on $T^{*}Q$. In Darboux' coordinates $(q^i,p_i)$ on $T^{*}Q$, the canonical 1-form $\theta_Q \in \Lambda^1(T^*Q)$ and the symplectic 2-form $\Omega_Q\in \Lambda^2(T^*Q)$ read 
\begin{equation} \label{can-forms}
\theta_Q=p_idq^i, \qquad \Omega_Q=dq^i\wedge dp_i,
\end{equation}
respectively. 
	
\subsubsection*{\rm \bf Special symplectic structures} 

Let $P$ be a symplectic manifold, that is, it is equipped with a non-degenerate closed 2-form called the \emph{symplectic 2-form}. Let, in particular, $P$ be equipped with an exact symplectic 2-form $\Omega=-d\theta$, where $\theta$ is called a \emph{potential 1-form}. Let, further, $P$ be the total space of a fiber bundle, say $\pi:P\to Q$, or $(P,\pi,Q)$ in short. A special symplectic structure, then, is a quintuple $(P,\pi,Q,\theta,\Theta)$, where $\Theta:P\to T^{\ast}Q$ is a fiber preserving symplectic diffeomorphism given by
\begin{equation} \label{chi-}
\langle \Theta(x),\pi_{\ast}X(q)\rangle =\langle\theta(x),X(x)\rangle
\end{equation}
for any vector field $X$ on $P$, and any point $x\in P$ with $\pi(x)=q$. Here, the pairing on the left hand side is the (natural) one between $T^*_qQ$ and $T_qQ$, whereas the pairing on the right hand side is the one between $T^*_{x}P$ and $T_{x}P$. In this case, the tuple $(P,\Omega)$ is called the \emph{underlying symplectic manifold of the special symplectic structure}. We refer the reader to \cite{LaSnTu75, SnTu72, Tu80} for further details on the special symplectic structures.

\subsection{Tulczyjew's triplet} \label{subsect-Tulczyjew-triplet}~
	
We shall next review the Tulczyjew's symplectic space and the construction of the Tulczyjew's triplet, in a very classical sense. 	
	
\subsubsection*{\rm\bf Two Operations} 

In order to be able to proceed to Tulczyjew's symplectic space, we shall need to recall two operations from $\Lambda(Q)$ to $\Lambda(TQ)$. Letting $T\tau_{Q}:TTQ\to TQ$ to be the tangent lift of the tangent bundle projection, the first operation is the one which assigns a $(k-1)$-form in $\Lambda(TQ)$ to each $k$-form in $\Lambda(Q)$. More explicitly,
\begin{equation}\label{i_T}
i_{T}:\Lambda^k  ( Q ) \to \Lambda^{k-1}  ( TQ), \qquad \Omega ^{k}\mapsto i_{T}\Omega ^{k},
\end{equation}
which is given by
\[
i_{T}\Omega ^{k}(X_1,\dots,X_{k-1} )=\Omega ^{k}\Big( 
\tau _{TQ}(X_1),T\tau _{Q}(X_1),\dots,T\tau _{Q}(X_{k-1}) \Big)
\]
for any vector fields $X_{1},\dots,X_{k-1}$ on $TQ$. The second operation, on the other hand, is a degree $0$ derivation, namely,
\begin{equation}\label{d-T}
d_{T}:\Lambda^k  ( Q ) \to \Lambda^{k}  ( TQ ),\qquad \Omega ^{k}\mapsto (i_{T}d+di_{T})\Omega ^{k},
\end{equation}	
where $d:\Lambda^k\to \Lambda^{k+1}$ refers to the deRham exterior derivative on the exterior algebra of the relevant manifold. For further details we refer the reader to \cite{Tu77, Tu80,tulczyjew2004homogenous, tulczyjew1999slow}.	

\subsubsection*{\rm \bf Tulczyjew's symplectic space}

We are now ready to review the Tulczyjew's symplectic space from \cite{Tu77}. The tangent bundle of a symplectic manifold $( P,\Omega ) $ is also a symplectic manifold given by $(TP,d_T\Omega )$. In particular, starting with the canonical symplectic manifold $(T^{\ast }Q,\Omega_{Q}:=-d\theta_{Q})$, \emph{Tulczyjew's symplectic space} $TT^*Q$ becomes a symplectic manifold equipped with the lifted symplectic 2-form $d_T\Omega_{Q}$ which admits two potential 1-forms
\begin{equation} \label{tet2}
\vartheta _{1}=-i_{T}\Omega_{Q}, \qquad  \vartheta _{2}=d_{T}\theta _{Q}=i_{T}d\theta_{Q}+di_{T}\theta _{Q}. 
\end{equation}
If $(q^i)$ is a coordinate chart on $Q$, we shall then make use of $(q^i,\dot{q}^j)$ on $TQ$ for the induced coordinates. Accordingly, the induced coordinates on $TT^*Q$ may be given by $(q^i,p_j,\dot{q}^k,\dot{p}_l)$, and the 1-forms \eqref{tet2} read 
\begin{equation}\label{TT*Q-1-forms}
\vartheta _{1}=-i_{T}\Omega_{Q}=\dot{p}_{i}dq^i-\dot{q}^i dp_i, \qquad 
\vartheta _{2}=d_{T}\theta _{Q}=\dot{p}_{i}dq^{i}+p_{i}d\dot{q}^{i},
\end{equation}
see, for instance, \cite{We83}. Furthermore, the symplectic 2-form on $TT^*Q$ appears to be
\begin{equation}\label{dT-Omega-Q}
d_T\Omega_Q=d\vartheta_1=d\vartheta_2=d\dot{p}_i\wedge dq^i   + dp_i\wedge d\dot{q}^i.
\end{equation}

\subsubsection*{\rm \bf The canonical involution on $TTQ$}

Along the lines of \cite[Sect. 5]{tulczyjew1999slow}, see also \cite{abraham1978foundations,GrUrRo10}, given a differential mapping $\g:\mathbb{R}^2\to{Q}$, $\g=\g(s,t)$, both
\[
\dot{\g}(s):=\left.\frac{\p \g(s,t)}{\p t}\right|_{t=0} \qquad \text{and}\qquad {\g}'(t):=\left.\frac{\p \g(s,t)}{\p s}\right|_{s=0}
\]
determine curves $\dot{\g},\g':\B{R}\to TQ$, so that 
\[
(\dot{\g})':= \left.\frac{d \dot{\g}(s)}{d s}\right|_{s=0} \in TTQ, \qquad \dot{(\g')}:= \left.\frac{d \g'(t)}{d t}\right|_{t=0} \in TTQ.
\]
Accordingly, the mapping given by
\begin{equation} \label{involution}
\kappa_Q:TT{Q}\to TT{Q}, \qquad \dot{\g}' \mapsto \dot{\g'}
\end{equation}  
is called the \emph{canonical involution} on $TTQ$. A quick inspection then reveals that the involution \eqref{involution} satisfies
\[
\tau_{T{Q}}\circ \kappa_Q =T\tau_{Q}, \qquad T\tau_{Q} \circ \kappa_Q =\tau_{T\mathcal{Q}}.
\]
In terms of the induced coordinates $(q^i,\dot{q}^j,{q}^{\prime k},\dot{q}^{\prime l })$ on the iterated tangent bundle $TTQ$, we have
\begin{equation}\label{local-tau-TQ-Q}
\tau_{TQ}(q^i,\dot{q}^j,{q}^{\prime k},\dot{q}^{\prime l })=(q^i,\dot{q}^j), \qquad T\tau_{Q}(q^i,\dot{q}^j,{q}^{\prime k},\dot{q}^{\prime l })=
(q^i, {q}^{\prime k}),
\end{equation}
and furthermore, the canonical involution \eqref{involution} is computed to be 
\[
\kappa _Q(q^i,\dot{q}^j,{q}^{\prime k},\dot{q}^{\prime \ell })=(q^i,{q}^{\prime k},\dot{q}^j,\dot{q}^{\prime \ell }).
\]

\subsubsection*{\rm \bf The pairing between $TT^*Q$ and $TTQ$}	

We shall now recall, also from \cite{tulczyjew1999slow}, a pairing between $TT^*Q$ and $TTQ$. To this end, given any $Z\in TT^*Q$ and any $W\in TTQ$ satisfying    $T\tau_{Q}(W)=T\tau^*_Q(Z)$, let $z(t) \in T^*Q$ be the curve with $\dot{z}(0)=Z$, and similarly let $w(t) \in T^*Q$ be the curve with $\dot{w}(0)=W$, so that $\tau_Q\circ w = \tau^*_Q\circ z$. Then, a pairing of bundles over $TQ$ may be formulated by
\begin{equation} \label{pairing-tilde}
\langle \bullet,\bullet \rangle^{\widetilde{}} :TT^\ast Q \times TTQ\to 
\mathbb{R}, \qquad \langle Z,W\rangle^{\widetilde{}} :=\left .\frac{d}{dt} \langle z(t),w(t) \rangle \right \vert_{t=0}.
\end{equation} 
Setting the (induced) coordinates on $TT^*Q$ as $(q^i,p_j,\dot{q}^k,\dot{p}_l)$, the coordinate expression of the pairing \eqref{pairing-tilde} may be given by 
\[
\big\langle (q^i,p_j,\dot{q}^k,\dot{p}_l),(q^i,{q}^{\prime j},\dot{q}^k,
\dot{q}^{\prime l })\big\rangle^{\widetilde{}} = p_i\dot{q}^{\prime i}+q^{\prime i}\dot{p}_i.
\]
	
We shall conclude the present subsection with two symplectomorphisms; $\a_Q:TT^*Q\to T^*TQ$, and $\Omega^\flat_{Q}:TT^*Q\to T^*T^*Q$. To this end, we shall assume that, being cotangent bundles, $T^*TQ$ and $T^*T^*Q$ equipped with the canonical symplectic forms $\Omega_{TQ}=-d\theta_{TQ}$ and $\Omega_{T^*Q}=-d\theta_{T^*Q}$, respectively.

\subsubsection*{\rm \bf The left Wing of the Tulczyjew's Triplet} 

The first symplectomorphism that we shall present is the morphism 
\begin{equation}\label{alpha}
\alpha_{Q}: TT^*Q \to T^*TQ, \qquad  \langle \alpha_{Q}(Z) ,W \rangle=-\langle Z, \kappa_{Q}(W) \rangle ^{\widetilde{}},  
\end{equation}
of vector bundles, which reads, in reduced coordinates
\[
\alpha _{Q}(q^i,p_j,\dot{q}^k,\dot{p}_l)=(q^i,\dot{q}^k, -\dot{p}_l, -p_j).
\]
We do note also that the pairing on the right hand side of \eqref{alpha} is the one in \eqref{pairing-tilde}, whereas the pairing of the left hand side is the canonical pairing between $T^*TQ$ and $TTQ$. 

A straightforward calculation reveals that $\alpha_Q^{\ast }\Omega _{TQ}=d_T\Omega_Q$, hence \eqref{alpha} is indeed a symplectomorphism. As a result, we arrive at a special symplectic structure
\begin{equation}\label{sss-1}
(TT^{\ast }Q,T\tau^*_{Q},TQ,\vartheta_{2},\alpha_{Q}) 
\end{equation}		
the underlying symplectic manifold of which being $(TT^*Q,d_T\Omega_Q)$.

\subsubsection*{\rm \bf The right Wing of the Tulczyjew's Triplet}  

The nondegeneracy of the canonical symplectic 2-form on $T^*Q$ leads to the existence of a (musical) diffeomorphism given by
\begin{equation} \label{beta}
\Omega^\flat_{Q}: TT^*Q\to T^*T^*Q, \qquad \Omega^\flat_Q(z)=\Omega_Q(z,\bullet),
\end{equation}
which may be presented in reduced coordinates as
\[
\Omega^\flat_{Q}(q^i,p_j,\dot{q}^k,\dot{p}_l)=(q^i,p_j,-\dot{p}_l,\dot{q}^k).
\]
Once again, a quick calculation yields $(\Omega^\flat_{Q})^{\ast }\Omega _{T^*Q}=d_T\Omega_Q$, and hence \eqref{beta} is a symplectomorphism. Accordingly, there is a special symplectic structure 
\begin{equation}\label{sss-2}
(TT^{\ast }Q,\tau _{T^{\ast }Q},T^{\ast }Q,\vartheta _{1},
\Omega _{Q}^{\flat }),
\end{equation}
over the (underlying) symplectic manifold $(TT^*Q,d_T\Omega_Q)$.
	
Referring the reader to \cite{tulczyjew1972hamiltonian, Tu77, Tu80, tulczyjew1999slow} for further details, let us finally record the following commutative diagram summarizing the entire discussion on the present subsection.    
\begin{equation} \label{TT}
\xymatrix{T^{\ast }TQ \ar[dr]_{\tau^*_{TQ}}&&TT^{\ast}Q\ar[dl]^{T\tau^*_{Q}}
\ar[rr]^{\Omega_{Q}^{\flat}} \ar[dr]_{\tau_{T^{\ast}Q}} \ar[ll]_{\alpha_{Q}}&&T^{\ast }T^{\ast}Q\ar[dl]^{\tau^* _{T^{\ast}Q}}
\\
&TQ\ar[dr]_{\tau_{Q}}&&T^{\ast}Q\ar[dl]^{\tau^*_{Q}} \\
&&Q}
\end{equation}

\subsection{Connection and curvature} \label{subsect-connection-curvature}~ 
	
In the present subsection we shall now review briefly the very basics of the theory connections, and curvatures.

Let us first recall that the kernel of the tangent lift $T\pi:TQ\to T\bar{Q}$ of the principal $G$-bundle $\pi:Q\to \bar{Q}$, $q\mapsto [q]$, determines the vertical subbundle $VQ$ of $TQ$. We note, on the other hand, that any element of the Lie algebra $\G{g}$ of the symmetry group $G$ generates a vertical vector field; an element of the space $\G{X}(Q)$ of sections of $TQ$ which takes values in the fibers of $VQ$, through
\begin{equation} \label{inf-gen-xi}
\xi_Q(q)=T_e\phi_q(\xi)=\left.\frac{d}{dt} \Big((\exp t\xi) \cdot q\Big) \right\vert_{t=0},
\end{equation}
where $\exp:\G{g}\to G$ is the exponential map, and $\xi \in \G{g}$. The map $\G{g}\mapsto \G{X}(Q)$, given by $\xi\mapsto\xi_Q$, is a Lie algebra homomorphism, and $\xi_Q \in \G{X}(Q)$ is called the fundamental (vertical) vector field associated to $\xi \in \G{g}$. 

Let us next recall, following \cite{cendra2003variational,KobaNomi-book-I} (see also \cite{n2001lagrangian,nakahara2003geometry}), a connection on the principal $G$-bundle $(Q,\pi,\bar{Q})$ is an (differentiable) assignment of a subspace $H_qQ \subseteq T_qQ$ to any $q\in Q$, so that
\begin{itemize}
\item[(i)] $T_pQ = V_pQ\oplus H_pQ$,
\item[(ii)] $H_{g\cdot q}Q = TL_g(H_qQ)$, for any $g\in G$.
\end{itemize}
Accordingly, in the presence of a connection, the tangent bundle decomposes into a Whitney sum
\[
TQ = VQ \oplus HQ
\]
of vertical and horizontal subbundles. Equivalently, a connection on the principal $G$-bundle $(Q,\pi,\bar{Q})$ may be viewed as a $\G{g}$-valued 1-form $A:TQ\to \G{g}$ satisfying
\begin{equation} \label{connection}
A \circ \xi_Q =\xi, \qquad A\circ T\phi_{g}=\Ad_{g}\circ A ,
\end{equation} 
where $\Ad:G\times \G{g}\to \G{g}$ is the adjoint representation of the group $G$ on its Lie algebra $\mathfrak{g}$. 

Let us note also that the tangent lift $T\pi:T_qQ\to T_{\pi(q)}\bar{Q}$ maps $H_qQ$ isomorphically onto $T_{\pi(q)}\bar{Q}$, for any $q\in Q$. Accordingly, following the notation in \cite[Sect. 2.2]{cendra2003variational}, given any $v_{\pi(q)} \in T_{\pi(q)}\bar{Q}$, we shall denote by $v_q^h \in T_qQ$ the unique horizontal vector satisfying 
\[
T_q\pi (v_q^h)=v_{\pi(q)} \in T_{\pi(q)}\bar{Q},
\]
and call it the \emph{horizontal lift} of the vector $v_{\pi(q)} \in T_{\pi(q)}\bar{Q}$. With a slight abuse of notation, we shall write
\begin{equation} \label{hor-lift-ope}
h:T\bar{Q}\to TQ, \qquad v_{\pi(q)} \mapsto v_q^h.
\end{equation}

Finally, the curvature of a connection $A:TQ\to \G{g}$ on the $G$-bundle $(Q,\pi,\bar{Q})$ is defined to be the $\G{g}$-valued 2-form given by (the Cartan structure equation)
\begin{equation}\label{curvature}
B(X_1,X_2)=dA(X_1,X_2)-[A(X_1),A(X_2)],
\end{equation} 
for any $X_1, X_2 \in \G{X}(Q)$.

\subsection{Trivialization and reduction of the (co)tangent bundles}\label{subsect-triv-reduct-tangent-cotangent}~ 

We shall next discuss the trivialization, and its reduction under a group action, of both the tangent and the cotangent bundles of the total space of a principal bundle. In order to develop some terminology, we shall begin with a quick detour on associated bundles.

\subsubsection*{\rm \bf Associated bundles} 

We shall now recall the construction of a vector bundle associated to a given principal bundle. In particular, we shall review the adjoint bundle and the coadjoint bundle constructions. 

To this end, let $\Phi:G\times V\to V$ denoted by $(g,v)\mapsto \Phi_g(v)$, represents a (differentiable) linear representation of $G$ on a vector space $V$. Then, the space $\widetilde{V}$ of orbits of $Q\times V$ with respect to the diagonal $G$-action
\begin{equation} \label{assoc-act}
G\times(Q\times V) \to (Q\times V), \qquad g\ast(q,v)= (g\cdot q, \Phi_g(v))
\end{equation}  
admits the structure of a vector bundle over $\bar{Q}$, given by
\begin{equation} \label{assoc-bundle}
\widetilde{V} \to \bar{Q}, \qquad   [q,v] \mapsto \pi(q),
\end{equation}
where $[q,v] \in \widetilde{V}$ represents the orbit of $(q,v)\in Q\times V$ with respect to \eqref{assoc-act}. 
Let us note also that any vector bundle over a manifold $M$, with fibers in an $n$-dimensional vector space, is an associated bundle of a $GL(n,\B{R})$-principal bundle called the \emph{frame bundle}.

In particular, taking the Lie algebra $\G{g}$ of the Lie group $G$ as the vector space, along with the adjoint representation of $G$ on $\G{g}$, we arrive to the associated vector bundle $\tilde{\G{g}}$, which is called the \emph{adjoint bundle} of the principal $G$-bundle $(Q,\pi,\bar{Q})$.

If, on the other extreme, one takes the linear dual $\G{g}^*$ as the vector space, and the coadjoint representation $\Ad^*:G\times \G{g}^* \to \G{g}^*$ given by
\begin{equation}\label{Ad*-operator}
\langle \Ad^*_g(\mu), \xi \rangle = \langle \mu, \Ad_{g^{-1}}(\xi)\rangle
\end{equation}
for any $\mu \in \G{g}^*$ and any $\xi \in \G{g}$, the corresponding associated bundle $\tilde{\G{g}}^*$ is called the \emph{coadjoint bundle} of $(Q,\pi,\bar{Q})$.

\subsubsection*{\rm \bf Trivialization and reduction of the tangent bundle} 

Given the principal $G$-bundle $(Q,\pi,\bar{Q})$, let us now consider the short exact sequence
\begin{equation}\label{sequence-bundles}
\xymatrix{
0 \ar[r] & VQ \ar[r] & TQ \ar[r] & \pi^\ast(T\bar{Q}) \ar[r] & 0
}
\end{equation}
of vector bundles over $Q$. In this framework, the presence of a connection $A:TQ\to \G{g}$, then, may be interpreted as the splitting of \eqref{sequence-bundles}, see for instance \cite{GrabKotoPonc11}. Accordingly, one may devise a diffeomorphism given by
\begin{align} \label{lambda}
\begin{split}
&\lambda_{T}:TQ \to Q\times_{\bar{Q}} T\bar{Q}\times\mathfrak{g},\qquad v \mapsto (\tau_Q(v),T\pi(v),A(v)), \\
&\lambda_{T}^{-1}: Q\times_{\bar{Q}} T\bar{Q}\times\mathfrak{g} \to TQ, \qquad (q,u,\xi) \mapsto u_q^h+\xi_{Q}(q),
\end{split}
\end{align}
for any $v\in TQ$, any $(q,u) \in Q\times_{\bar{Q}}T\bar{Q}$, and any $\xi\in\G{g}$. 

The diagonal action of $G$ on $Q\times \G{g}$, then, reduces \eqref{sequence-bundles} into the \emph{Atiyah sequence}
\begin{equation}\label{Atiyah-seq}
\xymatrix{
0 \ar[r] & \tilde{\G{g}} \ar[r] & G\backslash TQ \ar[r] & T\bar{Q} \ar[r] & 0
}
\end{equation}
of vector bundles over $\bar{Q}$, which also splits by the connection. Accordingly, the trivialization \eqref{lambda} reduces to that of
\begin{align} \label{bar-lambda}
\begin{split}
& \overline{TQ} \to  T\bar{Q}\times_{\bar{Q}} \tilde{\G{g}}, \qquad [v] \mapsto (T\pi(v), [\tau_Q(v),A(v)]), \\
& T\bar{Q}\times_{\bar{Q}} \tilde{\G{g}} \to \overline{TQ}, \qquad (u_{\pi(q)},[q,\xi]) \mapsto  [u^h_q +\xi_Q(q)], 
\end{split}
\end{align}
where $\overline{TQ}:=G\backslash TQ$.

\subsubsection*{\rm \bf Trivialization and reduction of the cotangent bundle} 

Next, by a slight abuse of language, dualizing \eqref{lambda} we arrive at a similar trivialization of the cotangent bundle, which may be given by
\begin{align}\label{lambda-T*}
\begin{split}
& \lambda_{T^*}:T^*Q \to Q\times_{\bar{Q}} T^*\bar{Q}\times \G{g}^*,\qquad z \mapsto (\tau^*_Q(z),h^*(z),\mathbf{J}_Q(z)),\\
&\lambda_{T^*}^{-1}:Q\times_{\bar{Q}} T^*\bar{Q}\times \G{g}^* \to T^*Q, \qquad (q,y,\mu) \mapsto T^*_q\pi (y)+A_q^*\mu,
\end{split}
\end{align}
for any $z\in T^*Q$, and any $(q,y) \in Q\times_{\bar{Q}} T^*\bar{Q}$, where
\begin{equation}\label{def-mom} 
\mathbf{J}_Q:T^*Q\to \mathfrak{g}^*, \qquad
\langle \mathbf{J}_Q(z),\xi\rangle :=\langle z,\xi_Q\rangle,
\end{equation}
is the \emph{moment map}, while $A_q^\ast:\G{g}^* \to T_q^*Q$ and $h^*:T_q^*Q\to T^*_{\pi(q)}\bar{Q}$ are the linear algebraic duals of the connection and the horizontal lift operator, respectively.

Finally, the reduction with respect to the diagonal action of $G$ on $Q\times \G{g}^*$ yields 
\begin{equation} \label{bar-lambda-T*}
\overline{\lambda_{T^*}}:\overline{T^*Q} \to T^*\bar{Q} \times_{\bar{Q}} \tilde{\G{g}}^*, \qquad [z] \mapsto (h^*(z), [\tau^*_Q(z),\mathbf{J}_Q(z_q)]),    
\end{equation}  
where $\overline{T^*Q}:=G\backslash T^*Q$ via the coadjoint lift of the $G$-action on $Q$.

\subsubsection*{\rm \bf Trivialized tangent - cotangent duality} 

Let us conclude with the manifestation of the natural pairing between the tangent bundle and the cotangent bundle in view of the trivializations \eqref{lambda} and \eqref{lambda-T*}. To this end, given any $T^*Q\ni z\simeq (q,y,\mu) \in Q \times T^*\bar{Q}\times \G{g}^*$, and any $TQ\ni v\simeq (q,u,\xi) \in Q \times TQ\times \G{g}$, we have
\[
T^*Q\times TQ\to \mathbb{R}, \qquad  \langle (q,y,\mu) ,(q,u,\xi) \rangle = \langle y,u \rangle + \langle \mu, \xi \rangle.
\]

\section{Trivializations and Reductions of the Iterated (Co)tangent Bundles}\label{Sec-TTQ-TT*Q}
	
In the present section we shall derive the trivializations and reductions of the iterated tangent and cotangent bundles. 

To this end, we shall first record the following terminology on the (tangent) group actions. Given a Lie group $G$, the structure of the group structure of the tangent group $TG$ may be (right) trivialized via
\[
tr_{TG}^R:TG\to \G{g} \rtimes G, \qquad v_g \mapsto (TR_{g^{-1}}v_g,g),  
\]
where $R_g:G\to G$ stands for the right translation of $G$, and the group operation of the latter is given by
\[
(\xi,g)(\eta,h) = (\xi+\Ad_g(\eta), gh)
\]
for any $\xi,\eta\in \G{g}$, and any $g,h\in G$. 

On the other hand, let us note that, the tangent mapping of the group action $\phi:G\times Q\to Q$ gives rise to the action
\begin{equation}\label{tan-action}
T\phi:TG\times TQ \to TQ, \qquad (\xi,g)\cdot v \mapsto T\phi_g(v)+\xi_Q(g\cdot \tau_Q(v)),
\end{equation} 
which also is free and proper, \cite{KolaMichSlov-book}. As for the reduction, on the other hand, we have
\begin{equation}\label{TG-reduction}
TG\backslash TQ \simeq (\mathfrak{g}\rtimes G)\backslash TQ \simeq \G{g}\backslash(G \backslash TQ) \simeq \G{g}\backslash(T\bar{Q} \times_{\bar{Q}} \tilde{\G{g}})\simeq T\bar{Q},
\end{equation} 
where the third identification is given by \cite[Lemma 2.4.2]{n2001lagrangian}. As a result, we obtain the $TG$-principal bundle $(TQ,T\pi,T\bar{Q})$. Furthermore, a straightforward calculation reveals that $A:TQ\to\G{g}$ being a connection on the $G$-bundle $(Q,\pi,\bar{Q})$, its tangent map $TA: TTQ\to \G{g} \rtimes \G{g}$ happens to satisfy the manifestations
\[
TA\circ (\xi,\eta)_{TQ}=(\xi,\eta), \qquad TA\circ T(T\phi)_{(\zeta,g)}=\Ad_{(\zeta,g)}\circ TA,
\]
of the requirements \eqref{connection} of a connection on the $TG$-bundle $(TQ,T\pi,T\bar{Q})$, where $(\xi,\eta)_{TQ} : TQ\to TTQ$ stands for the fundamental vertical vector field associated to $(\xi,\eta) \in \G{g}\rtimes \G{g}$, and $\G{g} \rtimes \G{g}$ is the Lie algebra of the tangent group $TG\simeq \G{g}\rtimes G$, whose structure is given by
\[
[(\xi_1,\eta_1),(\xi_2,\eta_2)] = ([\xi_1,\xi_2]+[\eta_1,\xi_2] - [\eta_2,\xi_1], [\eta_1,\eta_2])
\]
for any $\xi_1,\xi_2,\eta_1,\eta_2 \in \G{g}$. Finally, it follows from \cite[(2.14)]{EsenSutl17} that the adjoint action of the tangent group $TG$ on its Lie algebra $\G{g} \rtimes \G{g}$ is given by
\begin{equation}\label{Ad-action-on-Tg}
\Ad_{(\zeta,g)}:\G{g}\rtimes \G{g} \to \G{g}\rtimes \G{g} ,\qquad (\xi,\eta) \mapsto (\Ad_g\xi-[\Ad_g\eta,\zeta], \Ad_g\eta),
\end{equation}
for any $\xi, \eta, \zeta \in \G{g}$, and any $g\in G$.

Finally, we do note that the tangent lift $Th:TT\bar{Q} \to TTQ$ of \eqref{hor-lift-ope} 
works as the horizontal lift associated to the \emph{tangent connection} on $TTQ$.
	
We are now ready to proceed onto the trivializations and the reductions of the iterated tangent and  cotangent bundles.
	
\subsection{Trivialization and reduction of $TTQ$}\label{Sec-tri-TTQ}~

In this subsection we shall first derive a trivialization of $TTQ$, given the principal $G$-bundle $(Q,\pi,\bar{Q})$. Formulating the $G$-action on $TTQ$ in terms of this trivialization, we shall present explicitly the $G$-reduction of $TTQ$. Also in this subsection, we shall reformulate the canonical involution along the lines of the trivialization we obtain. Finally, we shall illustrate the reduction, under the $G$-action, of the canonical involution.

\subsubsection*{\rm \bf Trivialization of $TTQ$} 

To begin with, we record the straightforward identification
\[
T(Q\times_{\bar{Q}} T\bar{Q}\times \G{g})\simeq TQ\times_{T\bar{Q}}TT\bar{Q}\times T\G{g}.
\]
Then, it becomes a routine verification that the tangent lift of the trivialization (and its inverse) \eqref{lambda} may be formulated as
\begin{align}\label{tan-lambda}
\begin{split}
& T\lambda_{T}:TTQ \to  TQ\times_{T\bar{Q}} TT\bar{Q}\times T\G{g}, \qquad W\mapsto \big(T\tau_Q(W),TT\pi(W),TA(W)\big), \\
& T\lambda_{T}^{-1}:TQ\times_{T\bar{Q}} TT\bar{Q}\times T\G{g}\to TTQ,\qquad (v,U,\xi,\eta)\mapsto U^{Th}_v+(\xi,\eta)_{TQ}(v),
\end{split}
\end{align}
for any $W\in TTQ$, and any $(v,U,\xi,\eta)\in TQ\times_{T\bar{Q}}TT\bar{Q} \times T\G{g}$.

Next, in view of the decomposition $TQ \simeq VQ \oplus HQ$, along with the identification $T_q\pi:H_qQ \to T_{\pi(q)}\bar{Q}$, we have
\[
T(Q\times_{\bar{Q}} T\bar{Q}\times \G{g})\simeq TQ\times_{T\bar{Q}}TT\bar{Q}\times T\G{g} \simeq ((Q\times \G{g})\times_{\bar{Q}} T\bar{Q}) \times_{T\bar{Q}} TT\bar{Q}\times T\G{g} \simeq Q\times_{\bar{Q}}  TT\bar{Q}\times  T\G{g} \times \G{g},
\]
and hence obtain the following refinement of \eqref{tan-lambda}:
\begin{align} \label{Lambda-TT}
\begin{split}
 &\lambda_{TT}:TTQ \to Q\times_{\bar{Q}} TT\bar{Q}\times T\G{g}\times \G{g},\qquad\hspace{0.2 cm}
  W\mapsto (\tau_Q(T\tau_Q(W)),TT\pi(W),TA(W),A(T\tau_Q(W))), \\
 &\lambda^{-1}_{TT}:Q\times_{\bar{Q}} TT\bar{Q}\times T\G{g}\times \G{g} \to TTQ,\qquad (q,U,\xi,\eta,\zeta)  \mapsto U_{\tau_{T\bar{Q}}(U)_q^h+\zeta_Q(q)}^{Th}+(\xi,\eta)_{TQ}(\tau_{T\bar{Q}}(U)_q^h+\zeta_Q(q))
\end{split}
\end{align}
for any $W\in TTQ$, and any $(q,U,\xi,\eta,\zeta) \in Q\times_{\bar{Q}} TT\bar{Q}\times T\G{g}\times \G{g}$.

\subsubsection*{\rm \bf Trivialized canonical involution}

Before we move towards the reduction of $TTQ$ by the $G$-action, we shall now record the canonical involution \eqref{involution} on the trivialization \eqref{Lambda-TT} of $TTQ$. To this end let, once again, $\g:\B{R}^2\to Q$, $\g = \g(t,s)$ be such that $\g(0,0)=q \in Q$. Then the two curves on $TQ$ given by
\[
\dot{\g}(s):=\left.\frac{\p \g(s,t)}{\p t}\right|_{t=0} \qquad \text{and}\qquad {\g}'(t):=\left.\frac{\p \g(s,t)}{\p s}\right|_{s=0}
\]
may be trivialized, in view of \eqref{lambda}, into
\[
\dot{\Gamma}:\B{R}\to Q\times_{\bar{Q}}T\bar{Q}\times \G{g}\simeq TQ, \qquad s\mapsto (\g(0,s),x(s),\dot{x}(s),\xi(s))
\]
and
\[
\Gamma':\B{R}\to Q\times_{\bar{Q}}T\bar{Q}\times \G{g} \simeq TQ, \qquad t\mapsto (\g(t,0),x(t),x'(t),\zeta(t)),
\]
where 
\[
(x(s),\dot{x}(s)):=T\pi(\g(0,s),\dot{\g}(s)), \qquad (x(t),x'(t)):=T\pi(\g(t,0),\g'(t)),
\]
and
\[
\xi(s) := A(\g(0,s),\dot{\g}(s)), \qquad \zeta(t) := A(\g(t,0),\g'(t)).
\]
Accordingly, then, the canonical involution \eqref{involution} takes the form of
\[
\kappa_Q:TTQ\simeq T(Q\times_{\bar{Q}} T\bar{Q}\times \G{g})\to T(Q\times_{\bar{Q}} T\bar{Q}\times \G{g}) \simeq TTQ, \qquad (\g', x',\dot{x}',\xi') \mapsto (\dot{\g}, \dot{x},\dot{x'},\dot{\zeta}),
\]
where we suppressed the $0$'s, on which the derivatives are evaluated. 

Now, in an effort to recalibrate this last expression of the canonical involution according to \eqref{Lambda-TT}, we observe that
\begin{equation}\label{cal-can-inv}
 \dot{\zeta}-\xi'=\frac{d \zeta(t)}{dt}\Big|_{t=0}-\frac{d \xi(s)}{ds}\Big|_{s=0}=\frac{d A(\g(t,0), \g'(t))}{dt}\Big|_{t=0} - \frac{d A(\g(0,s),\dot{\g}(s))}{ds}\Big|_{s=0} ,
\end{equation}
where it follows from \cite[Sect. 3.1]{n2001lagrangian} that the latter term is the value at $t=0$ of the variation of $A(\g(t,0),\dot{\g}(t,0))$ corresponding to the variation $\g'(t)$ of the curve $\g(t,0) \in Q$. Along the lines of \cite[Lemma 3.1.1]{n2001lagrangian}, its vertical part; corresponding to the vertical variation ${\rm Ver}(\g(t,0)):= A(\g(t,0),\g'(t))\g(t,0)$ of the curve $\g(t,0) \in Q$, is
\[
\frac{d A(\g(t,0), \g'(t))}{dt}\Big|_{t=0} +[\zeta,\xi],
\] 
and it follows from \cite[Lemma 3.1.2]{n2001lagrangian} that its horizontal part is
\[
B_q(\g',\dot{\g}).
\]
Substituting these into \eqref{cal-can-inv} we arrive at
\begin{equation*}
\begin{split}
\kappa_Q:TTQ\simeq T(Q\times_{\bar{Q}} T\bar{Q}\times \G{g})\to T(Q\times_{\bar{Q}} T\bar{Q}\times \G{g}) \simeq TTQ, \qquad(\g', x',\dot{x}',\xi') \mapsto (\dot{\g}, \dot{x},\dot{x'},\xi'+B_q(\dot{\g},\g')+[\xi,\zeta]).
\end{split}
\end{equation*}

As a result, the canonical involution \eqref{involution} trivializes, in the level of \eqref{Lambda-TT}, into
\begin{align}\label{kappa-Q-trv}
\begin{split}
&\widehat{\kappa_{Q}}:Q\times_{\bar{Q}} TT\bar{Q}\times T\G{g}\times \G{g} \to Q\times_{\bar{Q}} TT\bar{Q}\times T\G{g}\times \G{g}, \\ 
&(q,U,\xi,\eta,\zeta) \mapsto (q,\kappa_{\bar{Q}}(U),\zeta,\eta+B((\tau_{T\bar{Q}}(U))_q^h,(T\tau_{\bar{Q}}(U))^h_q)+[\xi,\zeta],\xi),		
\end{split}
\end{align}
with $U=(x,\dot{x},x',\dot{x}') \in TTQ$, for which
\[
\tau_{T\bar{Q}}(U)=\tau_{T\bar{Q}}(x,\dot{x}, x', \dot{x}')=(x,\dot{x}), \qquad T\tau_{\bar{Q}}(U)=T\tau_{\bar{Q}}(x,\dot{x}, x', \dot{x}')=(x,  x').
\]
Let us note also that in the present framework we have
\[
\widehat{\tau_{TQ}}:Q\times_{\bar{Q}} TT\bar{Q}\times T\G{g}\times \G{g} \simeq TTQ \to TQ \simeq Q\times_{\bar{Q}} T\bar{Q}\times \G{g}, \qquad (q,U,\xi,\eta,\zeta) \mapsto (q,\tau_{T\bar{Q}}(U),\xi)
\]
and
\[
\widehat{T\tau_Q}:Q\times_{\bar{Q}} TT\bar{Q}\times T\G{g}\times \G{g} \simeq TTQ \to TQ \simeq Q\times_{\bar{Q}} T\bar{Q}\times \G{g}, \qquad (q,U,\xi,\eta,\zeta) \mapsto (q,T\tau_{\bar{Q}}(U),\zeta).
\]

\subsubsection*{\rm \bf Reduction of $TTQ$}

We shall now proceed towards the reduction of $TTQ$ under the Lie group action
\begin{equation}\label{TTAct}
G\times TTQ\to TTQ, \qquad g\cdot W:=TT\phi_g(W),
\end{equation} 
which, in terms of the trivialization \eqref{Lambda-TT}, may also be given by
\begin{align} \label{red-TTAct}
\begin{split}
& G \times (Q\times_{\bar{Q}} TT\bar{Q}\times T\G{g}\times \G{g}) \to Q\times_{\bar{Q}} TT\bar{Q}\times T\G{g}\times \G{g},\\
&(g, (q,U,\xi,\eta,\zeta)) \mapsto g\cdot(q,U,\xi,\eta,\zeta) := (g\cdot q, U, \Ad_{g}\xi,\Ad_{g}\eta,\Ad_{g}\zeta).
\end{split}
\end{align}
Accordingly, it follows at once that
\[
G\backslash TTQ \simeq TT\bar{Q} \times_{\bar{Q}}  \widetilde{\G{G}} ,
\]
where $\widetilde{\G{G}}:= G\backslash (Q\times \G{g}\times \G{g}\times \G{g}) \simeq \widetilde{\G{g}} \times_{\bar{Q}} \widetilde{\G{g}} \times_{\bar{Q}}\widetilde{\G{g}}$ through
\[
G\times (Q\times \G{g}\times \G{g}\times \G{g})\to Q\times \G{g}\times \G{g}\times \G{g}, \qquad (g, (q,\xi,\eta,\zeta))\mapsto( g\cdot q,\Ad_g\xi,\Ad_g\eta,\Ad_g\zeta),
\]
and that
\begin{align}
\begin{split}
&\overline{\kappa_Q}:TT\bar{Q}\times_{\bar{Q}} \widetilde{\G{G}}\to  TT\bar{Q}\times_{\bar{Q}} \widetilde{\G{G}},\\
&(U,[q,\xi,\eta,\zeta]) \mapsto (\kappa_{\bar{Q}}(U),[q,\zeta,\eta +B((\tau_{T\bar{Q}}(U))^h_q,(T\tau_{\bar{Q}}(U))^h_q)+[\xi,\zeta],\xi]),
\end{split}
\end{align}
which serves as the reduction of the canonical inclusion, for any $U \in TT\bar{Q}$, and any $[q,\xi,\eta,\zeta] \in \widetilde{\G{G}}$. Let us also note, in this case, that
\begin{align*}
& \overline{\tau_{TQ}}:TT\bar{Q}\times_{\bar{Q}} \widetilde{\G{G}} \to  T\bar{Q}\times_{\bar{Q}} \tilde{\G{g}}, \qquad (U,[q,\xi,\eta,\zeta]) \mapsto (\tau_{T\bar{Q}}(U),[q,\xi]),\\
&\overline{T\tau_{Q}}:TT\bar{Q}\times_{\bar{Q}} \widetilde{\G{G}} \to  T\bar{Q}\times_{\bar{Q}} \tilde{\G{g}} , \qquad (U,[q,\zeta,\xi,\eta]) \mapsto (T\tau_{\bar{Q}}(U),[q,\zeta]).
\end{align*}

\begin{remark}
Let us conclude with the reduction of $TTQ$, with respect to the tangent group action. Along with the trivialization \eqref{tan-lambda} of $TTQ$, the tangent group action may be given by
\begin{align}\label{TG-action-on-TTQ-trivialized}
\begin{split}
TG \times (TQ\times_{T\bar{Q}} TT\bar{Q}\times T\G{g}) \to TQ\times_{T\bar{Q}} TT\bar{Q}\times T\G{g}, \qquad((\xi,g), (v,U,\zeta,\eta))\mapsto ((\xi,g)\cdot v,U,\Ad_{(\xi,g)}(\zeta,\eta)), 
\end{split}
\end{align}
where the former component is the tangent group action \eqref{tan-action}, and the latter component refers to the adjoint action \eqref{Ad-action-on-Tg}. As a result, in view of \eqref{TG-reduction}, we arrive at
\begin{align*}
&\overline{T\lambda_T}:TG\backslash TTQ \to TG\backslash(TQ\times_{T\bar{Q}} TT\bar{Q}\times T\G{g})\simeq TT\bar{Q}\times_{T\bar{Q}} \widetilde{\G{H}},\qquad [W]_{TG} \mapsto (TT\pi(W),[v,TA(W)]_{TG}), \\
&\overline{T\lambda_T}^{-1}: TT\bar{Q}\times_{T\bar{Q}} \widetilde{\G{H}} \to TG\backslash TTQ, \qquad (U,[v,(\xi,\eta)]_{TG})\mapsto [U_{v}^{Th}+(\xi,\eta)_{TQ}(v)]_{TG}.
\end{align*} 
for any $[W]_{TG} \in TG\backslash TTQ$, any $U\in T_{(\pi(q),T\pi(v))}T\bar{Q}$, and any $(\xi,\eta) \in T\G{g}$, where $\widetilde{\G{H}}:=TG\backslash (TQ\times T\G{g})$ through the diagonal action of $TG$ on $TQ\times T\G{g}$. Moreover, there is an associated bundle structure given by
\[
\widetilde{\G{H}}:=TG\backslash (TQ\times T\G{g}) \to T\bar{Q}\simeq TG\backslash TQ, \qquad [v,\xi,\eta]_{TG} \mapsto [v]_{TG}.
\]
\qed
\end{remark}

\subsection{Trivialization and reduction of $TT^*Q$}\label{Sec-tri-TT*Q}~

We shall now consider the trivialization, and then the reduction of $TT^*Q$. Upon formulating the trivialization of $TT^*Q$, in terms of \eqref{lambda-T*}, we shall introduce the (trivialized) $G$-action on $TT^*Q$, via which we shall arrive at the $G$-reduction of $TT^*Q$. 

\subsubsection*{\rm \bf Trivialization of $TT^*Q$} 

Differentiating \eqref{lambda-T*}, we attain a natural trivialization of $TT^*Q$ as     
\begin{align} \label{T-lambda-T*}
\begin{split}
&T\lambda_{T^*}: TT^*Q \to TQ\times_{T\bar{Q}} TT^*\bar{Q}\times T\G{g}^*, \qquad Z \mapsto (T\tau^*_Q(Z),Th^*(Z),T\mathbf{J}_Q(Z)), \\
& T\lambda_{T^*}^{-1}:TQ\times_{T\bar{Q}} TT^*\bar{Q}\times T\G{g}^* \to TT^*Q,\qquad (v,Y,\mu,\nu)\mapsto TT^*\pi (Y)+TA^*(\mu,\nu),
\end{split}
\end{align} 
for any $Z\in TT^*Q$, and any $(v,Y,\mu,\nu)\in TQ\times_{T\bar{Q}}  TT^*\bar{Q}\times T\G{g}^*$, where $Th^*:TT^*Q\to TT^*\bar{Q}$ is the tangent lift of the linear algebraic dual of \eqref{hor-lift-ope}, $TA^*:T\G{g}^\ast \to TT^*Q$ is the tangent lift of the dual of the connection 1-form on $TQ$, and $T{\rm \bf J}_Q:TT^*Q \to T\G{g}^*$ is the tangent lift of the moment map \eqref{def-mom}.

Similar to the trivialization \eqref{Lambda-TT} of $TTQ$, employing the identification \eqref{lambda}, we next obtain
\begin{align}\label{Lambda-TT*}
\begin{split}
&\lambda_{TT^*}: TT^*Q\to Q\times_{\bar{Q}} TT^*\bar{Q}\times T\G{g}^*\times \G{g}, \hspace{0.5 cm} Z \mapsto (\tau_Q( T\tau^*_Q(Z)),Th^*(Z), T\mathbf{J}_Q(Z),A(T\tau^*_Q(Z)) ),\\
& \lambda_{TT^*}^{-1}: Q\times_{\bar{Q}} TT^*\bar{Q}\times T\G{g}^* \times \G{g}\to TT^*Q,\qquad(q,Y,\mu,\nu,\zeta) \mapsto T_{T_q^*\pi(\tau_{T^*\bar{Q}}(Y))}T^*\pi (Y)+TA^*(\mu,\nu) + {\rm \bf J}^*_{T^*Q}(\zeta)
\end{split}
\end{align}
for any $Z \in TT^*Q$, and any $(q,Y,\mu,\nu,\zeta) \in Q\times_{\bar{Q}} TT^*\bar{Q} \times T\G{g}^*\times \G{g}$, where ${\rm \bf J}^*_{T^*Q}:\G{g}\to TT^*Q$ is the dualization of ${\rm \bf J}_{T^*Q}:T^*T^*Q\to \G{g}^*$.

In this language, the maps $T\tau^*_Q:TT^*Q\to TQ$ and $\tau_{T^*Q}:TT^*Q\to T^*Q$ take the form of    
\begin{align}\label{triv-projs}
\begin{split}
&\widehat{T\tau^*_Q}:Q\times_{\bar{Q}} TT^*\bar{Q}\times T\G{g}^* \times \G{g}\to Q\times_{\bar{Q}} T\bar{Q} \times \G{g}, \hspace{1.1 cm} (q,Y,\mu,\nu,\zeta)\mapsto (q,T\tau^*_{\bar{Q}}(Y),\zeta), \\
& \widehat{\tau_{T^*Q}}:Q\times_{\bar{Q}} TT^*\bar{Q}\times T\G{g}^*\times \G{g}\to Q\times_{\bar{Q}} T^*\bar{Q} \times \G{g}^*, \qquad (q,Y,\mu,\nu,\zeta)\mapsto (q,\tau_{T^*\bar{Q}}(Y),\mu),
\end{split}
\end{align}
for any $(q,Y,\mu,\nu,\zeta) \in Q\times_{\bar{Q}} TT^*\bar{Q} \times T\G{g}^*\times \G{g}$.

Finally, given any $W\in TTQ$, and any $Z\in TT^*Q$, setting 
\[
\left\langle \lambda_{TT^*}(Z),\lambda_{TT}(W) \right\rangle^{\widehat{}}:=\langle Z, W \rangle^{\widetilde{}},
\]
we obtain the trivialization 
\begin{equation}\label{pairing-trv-symbol-126}
\langle \bullet, \bullet \rangle^{\widehat{}}:\left(Q\times_{\bar{Q}} TT^*\bar{Q} \times T\G{g}^* \times \G{g}\right)\times (Q\times_{\bar{Q}} TT\bar{Q}\times T\G{g}\times \G{g}) \to \B{R}
\end{equation}
of the pairing \eqref{pairing-tilde} in the form of
\[
\left\langle (q,Y,\mu,\nu,\zeta);(q,U,\xi,\eta,\zeta)   \right\rangle^{\widehat{}}= \langle Y,U \rangle ^{\widetilde{}}+\frac{d}{ds}\Big|_{s=0}\langle \mu+s\nu, \xi+s\eta    \rangle=\langle Y,U \rangle ^{\widetilde{}}+ \left\langle \nu, \xi \right\rangle+ \left\langle \mu, \eta \right\rangle.
\]

\subsubsection*{\rm \bf Reduction of $TT^*Q$} 

We now proceed, along the lines of \cite{yoshimura2009dirac}, to the reduction of $TT^*Q$ by the $G$-action
\begin{equation} \label{act-TT*Q}
G\times TT^*Q \to TT^*Q, \qquad g\cdot Z:=TT^*\phi_{g^{-1}} (Z),
\end{equation}
which trivializes into
\begin{equation} \label{G-act-triTT*Q}
\begin{split}
&G\times (Q\times_{\bar{Q}} TT^*\bar{Q}\times T\G{g}^*\times \G{g})\to Q\times_{\bar{Q}} TT^*\bar{Q}\times T\G{g}^*\times \G{g}, \\
& (g, (q,Y,\mu,\nu ,\zeta))\mapsto g\cdot (q,Y,\mu,\nu ,\zeta):= (g\cdot q,Y,\Ad_g^*\mu,\Ad_g^*\nu,\Ad_g\zeta).
\end{split}
\end{equation}
Accordingly, we have
\[
G\backslash TT^*Q \simeq TT^*\bar{Q} \times_{\bar{Q}} \widetilde{\G{K}},
\]
where $\widetilde{\G{K}}:= G\backslash (Q\times \G{g}^*\times \G{g}^*\times \G{g}) \simeq \widetilde{\G{g}}^* \times_{\bar{Q}} \widetilde{\G{g}}^* \times_{\bar{Q}}\widetilde{\G{g}}$ via
\[
G\times (Q\times \G{g}^*\times \G{g}^*\times \G{g})\to Q\times \G{g}^*\times \G{g}^*\times \G{g}, \qquad (g, (q,\mu,\nu,\zeta))\mapsto( g\cdot q,\Ad^*_g\mu,\Ad^*_g\nu,\Ad_g\zeta).
\]
Finally, the reductions of \eqref{triv-projs} then appear as
\begin{align*}
&\overline{T\tau^*_Q}:TT^*\bar{Q}\times_{\bar{Q}} \widetilde{\G{K}} \to  T\bar{Q}\times_{\bar{Q}} \tilde{\G{g}}, \hspace{1.1 cm} (Y,[q,\mu,\nu,\zeta]) \mapsto (T\tau^*_{\bar{Q}}(Y),[q,\zeta]),\\
&\overline{\tau_{T^*Q}}:TT^*\bar{Q}\times_{\bar{Q}} \widetilde{\G{K}} \to T^*\bar{Q}\times_{\bar{Q}} \tilde{\G{g}}^*, \qquad (Y,[q,\mu,\nu,\zeta]) \mapsto (\tau_{T^*\bar{Q}}(Y),[q,\mu]).
\end{align*}

\subsection{Trivialization and reduction of $T^*TQ$}\label{Sec-tri-T*TQ}~

In the present subsection we shall now study the case of $T^*TQ$. As usual, we shall first derive a trivialization of $T^*TQ$ out of \eqref{lambda-T*}. Then, along the lines of the $G$-action on $T^*TQ$, which is in fact the one that follows from the $G$-action on $T^*Q$ and on $TQ$, we shall present the $G$-reduction of $T^*TQ$.

\subsubsection*{\rm \bf Trivialization of  $T^*TQ$}

Replacing the principal $G$-bundle $(Q,\pi,\bar{Q})$ with the principal $TG$-bundle $(TQ,T\pi,T\bar{Q})$ in \eqref{lambda-T*}, we achieve at once
\[
T^*TQ \simeq TQ\times_{T\bar{Q}} T^*T\bar{Q}\times (T\G{g})^*,
\]
the composition of which with $TQ\simeq Q\times_{\bar{Q}} T\bar{Q}\times \G{g}$ then quickly yields
\begin{align}\label{lambda-T*T}
\begin{split}
& \lambda_{T^*T}:T^*TQ \to Q\times_{\bar{Q}} T^*T\bar{Q} \times (T\G{g})^* \times \G{g}, \hspace{0.3 cm} \Upsilon \mapsto (\tau_Q(\tau^*_{TQ}(\Upsilon)), T^*h(\Upsilon), {\rm \bf J}_{TQ}(\Upsilon), A(\tau^*_{TQ}(\Upsilon))), \\
& \lambda_{T^*T}^{-1}: Q\times_{\bar{Q}} T^*T\bar{Q} \times (T\G{g})^* \times \G{g} \to T^*TQ, \qquad(q,K,\mu,\nu,\zeta) \mapsto T_{\tau^*_{T\bar{Q}}(K)_q^h+\zeta_Q(q)}^*(T\pi)(K) + (TA)^*_{\tau^*_{T\bar{Q}}(K)_q^h+\zeta_Q(q)}(\mu,\nu),
\end{split}
\end{align}
where $T^*h:T^*TQ\to T^*T\bar{Q}$ is the cotangent lift of \eqref{hor-lift-ope}, $TA:TTQ\to T\G{g}$ is the connection 1-form on $(TTQ,T\pi,T\bar{Q})$, with $(TA)^*_{\tau^*_{T\bar{Q}}(K)_q^h+\zeta_Q(q)}:(T\G{g})^*\to T^*_{\tau^*_{T\bar{Q}}(K)_q^h+\zeta_Q(q)}TQ$ being its linear algebraic dual, and ${\rm \bf J}_{TQ}:T^*TQ\to (T\G{g})^*$ is the mapping given by
\[
\langle {\rm \bf J}_{TQ}(\Upsilon), (\xi,\eta)\rangle = \langle \Upsilon, (\xi,\eta)_{TQ}(\tau^*_{TQ}(\Upsilon))\rangle
\]
for any $\Upsilon \in T^*TQ$, and any $(q,K,\mu,\nu,\zeta) \in Q\times_{\bar{Q}} T^*T\bar{Q} \times (T\G{g})^* \times \G{g}$.

Let us record also that the trivialized projection $\tau^*_{TQ}:T^*TQ\to TQ$ then, in this framework, reads
\begin{equation}\label{widehat-pi-TQ}
\widehat{\tau^*_{TQ}}:  Q\times_{\bar{Q}} T^*T\bar{Q}\times (T\G{g})^*\times \G{g} \simeq T^*TQ\to TQ\simeq Q\times_{\bar{Q}} T\bar{Q}\times \G{g}, \hspace{0.5 cm}(q,K,\mu,\nu,\zeta) \mapsto (q,\tau^*_{T\bar{Q}}(K),\zeta),
\end{equation} 
for any $(q,K,\mu,\nu,\zeta) \in Q\times_{\bar{Q}} T^*T\bar{Q}\times (T\G{g})^*\times \G{g}$.

\begin{remark}
Let us note that this trivialization may also be accomplished by the dualization of the fibers of $(TTQ,\tau_{TQ},TQ)$ in view of the trivialization 
\begin{align}\label{pairing-T*TQ-TTQ}
\begin{split}
& \langle \bullet, \bullet \rangle:(Q\times_{\bar{Q}} T^*T\bar{Q} \times (T\G{g})^* \times \G{g})\times (Q\times_{\bar{Q}} TT\bar{Q}\times T\G{g}\times \G{g}) \to \B{R}, \\
& \left\langle (q,K,\mu,\nu,\zeta);(q,U,\xi,\eta,\zeta)   \right\rangle=\langle K,U \rangle + \left\langle \mu, \xi \right\rangle+ \left\langle \nu, \eta \right\rangle.
\end{split}
\end{align}
of the natural pairing between $T^*TQ$ and $TTQ$. 
\qed
\end{remark}

\subsubsection*{\rm \bf Reduction of $T^*TQ$}

We shall now present the reduction of $T^*TQ$ with respect to the cotangent lift of the $G$-action on $TQ$, namely,
\begin{equation}\label{action-T^*T}
G\times T^*TQ \to T^*TQ,\qquad g\cdot \Upsilon:=T^*T\phi_{g^{-1}}(\Upsilon), 
\end{equation}
for any $g\in G$, and any $\Upsilon \in T^*TQ$. To this end, we first note the expression 
\begin{align*}
& G\times (Q\times_{\bar{Q}} T^*T\bar{Q}\times (T\G{g})^*\times \G{g}) \to Q\times_{\bar{Q}} T^*T\bar{Q}\times (T\G{g})^*\times \G{g},\\
& (g, (q,K,\mu,\nu,\zeta))\mapsto  g\cdot (q,K,\mu,\nu,\zeta)=(g\cdot q,K,\Ad_g^*\mu,\Ad_g^*\nu,\Ad_g\zeta), 
\end{align*}
of \eqref{action-T^*T} in terms of the trivialization \eqref{lambda-T*T}. It, then, follows at once that
\[
G\backslash T^*TQ \simeq T^*T\bar{Q} \times_{\bar{Q}} \widetilde{\G{K}}.
\]
Accordingly, the (trivialized) projection \eqref{widehat-pi-TQ} above reduces to
\begin{equation}\label{red-pi-TQ} 
\overline{\tau^*_{TQ}}:T^*T\bar{Q}\times_{\bar{Q}} \widetilde{\G{K}}\to T\bar{Q}\times_{\bar{Q}} \tilde{\G{g}}, \qquad (K,[q,\mu,\nu,\zeta]) \mapsto (\tau^*_{T\bar{Q}}(K),[q,\zeta]). 
\end{equation} 
On the other hand, 
\begin{align}\label{pair-T*TQ-TTQ}
\begin{split}
&\langle \bullet, \bullet\rangle:(T^*T\bar{Q}\times_{\bar{Q}} \widetilde{\G{K}})\times(TT\bar{Q}\times_{\bar{Q}} \widetilde{\G{K}})\to \B{R}  \\
& \big\langle (K,[q,\mu,\nu,\zeta]), (U,[q,\xi,\eta,\zeta]) \big\rangle=\left\langle K,U \right\rangle+\left\langle \mu,\xi\right\rangle +\left\langle \nu,\eta \right\rangle 
\end{split}
\end{align}
determines a pairing on the reduced spaces, which then may be considered as the reduction of the (trivialized) pairing \eqref{pairing-T*TQ-TTQ}.

\subsection{Trivialization and reduction of $T^*T^*Q$}\label{Sec-tri-T*T*Q}~

We shall now conclude the present section with the trivialization, and then the reduction, of the iterated cotangent bundle $T^*T^*Q$. 

\subsubsection*{\rm \bf Trivialization of $T^*T^*Q$}

In order to obtain a trivialization of $T^*T^*Q$, we consider the fiberwise dualization of the (trivialized) bundle $(Q\times_{\bar{Q}}TT^*\bar{Q}\times T\G{g}^*\times \G{g}\simeq TT^*Q,\widehat{\tau_{T^*Q}},T^*Q \simeq Q\times_{\bar{Q}} T^*\bar{Q}\times \G{g}^*)$. In view of the explicit expression of the (trivialized) bundle projection $\widehat{\tau_{T^*Q}}:Q\times_{\bar{Q}}TT^*\bar{Q}\times T\G{g}^*\times \G{g} \to Q\times_{\bar{Q}} T^*\bar{Q}\times \G{g}^*$ in \eqref{triv-projs}, we have the (trivialized) bundle projection
\begin{equation}\label{widehat-pi-T*Q}
\widehat{\tau^*_{T^*Q}}:Q\times_{\bar{Q}} T^*T^*\bar{Q}\times T^*\G{g}^* \times \G{g}^* \to Q\times_{\bar{Q}} T^*\bar{Q}\times \G{g}^*,\qquad (q,L,\mu,\eta,\rho)\mapsto (q,\tau^*_{T^*\bar{Q}}(L),\mu).
\end{equation}
The explicit identification with $T^*T^*Q$, on the other hand, may indeed be given by
\begin{align}\label{lambda-T*T*}
\begin{split}
& \lambda_{T^*T^*}:T^*T^*Q \to Q\times_{\bar{Q}} T^*T^*\bar{Q}\times T^*\G{g}^* \times \G{g}^* , \qquad \Xi\mapsto (\tau^*_Q(\tau^*_{T^*Q}(\Xi)), T^*T^*\pi(\Xi), T^*A^*(\Xi), {\rm \bf J}_Q(\tau^*_{T^*Q}(\Xi))), \\
& \lambda_{T^*T^*}^{-1}: Q\times_{\bar{Q}} T^*T^*\bar{Q}\times T^*\G{g}^* \times \G{g}^* \to T^*T^*Q , \\
&\hspace{4 cm}  (q,L,\mu,\eta,\rho) \mapsto T^*_{T^*_q\pi(\tau^*_{T^*\bar{Q}}(L))+A_q^*(\rho)}h^*(L) + T^*_{T^*_q\pi(\tau^*_{T^*\bar{Q}}(L))+A_q^*(\rho)}{\rm \bf J}_Q(\mu,\eta),
\end{split}
\end{align}
for any $\Xi \in T^*T^*Q$, and any $(q,L,\mu,\eta,\rho) \in Q\times_{\bar{Q}} T^*T^*\bar{Q}\times T^*\G{g}^* \times \G{g}^*$.

\subsubsection*{\rm \bf Reduction of $T^*T^*Q$}

Let us finally note the reduction of the bundle $T^*T^*Q$ under the group action given by the double cotangent lift of the group action on the base manifold; namely,
\begin{equation}\label{action-T*T*}
G\times T^*T^*Q \to T^*T^*Q,\qquad (g,\Xi)\mapsto g\cdot \Xi:=T^*T^*\phi_{g^{-1}}(\Xi),
\end{equation}
for any $g\in G$, and any $\Xi\in T^*T^*Q$. On the level of the trivialization \eqref{lambda-T*T*} then, the group action appears to be
\begin{align*}
& G\times (Q\times_{\bar{Q}} T^*T^*\bar{Q}\times T^*\G{g}^*\times \G{g}^*) \to (Q\times_{\bar{Q}} T^*T^*\bar{Q}\times T^*\G{g}^*\times \G{g}^*),\\
& (g, (q,L,\mu,\eta,\rho)) \mapsto  g\cdot (q,L,\mu,\eta,\rho)=(g\cdot q,L, \Ad^*_g\mu, \Ad_g\eta,\Ad^*_g\rho),
\end{align*}
for any $g\in G$, and any $(q,L,\mu,\eta,\rho) \in Q\times_{\bar{Q}} T^*T^*\bar{Q}\times T^*\G{g}^*\times \G{g}^*$. As a result, we obtain at once
\[
G\backslash T^*T^*Q \simeq T^*T^*\bar{Q}\times \widetilde{\G{L}},
\]
where $\widetilde{\G{L}} := G\backslash (Q \times \G{g}^*\times \G{g}\times \G{g}^*) \simeq \widetilde{\G{g}}^*\times_{\bar{Q}} \widetilde{\G{g}}\times_{\bar{Q}} \widetilde{\G{g}}^*$ via the diagonal $G$-action, that is,
\[
G\times (Q \times \G{g}^*\times \G{g}\times \G{g}^*) \to Q \times \G{g}^*\times \G{g}\times \G{g}^*, \qquad g\cdot (q,\mu,\eta,\rho) := (g\cdot q, \Ad^*_g\mu, \Ad_g\eta,\Ad^*_g\rho).
\]
The bundle projection, on the other hand, then reduces to that of
\begin{equation}\label{red-trv-pi-T*Q}
\overline{\tau^*_{T^*Q}} : T^*T^*\bar{Q} \times_{\bar{Q}} \widetilde{\G{L}}\to T^*\bar{Q}\times_{\bar{Q}} \widetilde{\G{g}}^*,\qquad (L,[q,\mu,\eta,\rho])\mapsto (\tau^*_{T^*\bar{Q}}(L),[q,\mu]),
\end{equation}
while the natural pairing between the cotangent bundle $T^*T^*G$ and the tangent bundle $TT^*Q$ takes the form of
\begin{align}\label{pair-red-T*T*Q-TT*Q}
\begin{split}
&\langle \bullet,\bullet  \rangle :(T^*T^*\bar{Q} \times_{\bar{Q}} \widetilde{\G{L}}) \times (TT^*\bar{Q}\times_{\bar{Q}}\widetilde{\G{K}}) \to \B{R}, \\
&  \langle (L,[q,\mu,\eta,\rho]), (Y,[q,\mu,\nu,\zeta]) \rangle=\langle  L,Y \rangle +\langle \nu, \eta  \rangle +\langle  \rho, \zeta \rangle ,
\end{split}
\end{align}
for any $(L,[q,\mu,\eta,\rho]) \in T^*T^*\bar{Q} \times_{\bar{Q}} \widetilde{\G{L}}$, and any $(Y,[q,\mu,\nu,\zeta]) \in TT^*\bar{Q}\times_{\bar{Q}}\widetilde{\G{K}}$.
\section{Trivialization and Reduction of Tulczyjew's Triplet}\label{Red-Trv-ssp}

\subsection{Trivializations and reductions of the canonical forms}\label{subsect-canonical-forms}~

In this first subsection we shall present both the canonical 1-form and the symplectic 2-form in \eqref{can-forms} along the lines of the trivializations (and then the reductions) of the cotangent bundles $T^*Q$ and $T^*T^*Q$.
\subsubsection*{\rm \bf Trivialization and reduction of the canonical 1-form}~

Let us begin with the canonical 1-form $\t_Q\in \Lambda^1(T^*Q)$. Regarding it as a section of the cotangent bundle $\tau^*_{T^*Q}:T^*T^*Q\to T^*Q$, the defining relation \eqref{can-Lio} of the canonical 1-form may be incarnated as
\begin{align}\label{LHS-cal-theta-Q}
\begin{split}
&\langle \widehat{\theta_Q}(q,y,\mu),(q,Y,\mu,\nu,\zeta)\rangle=\langle \widehat{\tau_{T^*Q}}(q,Y,\mu,\nu,\zeta),\widehat{T\tau^*_Q}(q,Y,\mu,\nu,\zeta) \rangle =\\
&\langle (q,\tau_{T^*\bar{Q}}(Y),\mu),(q,T\tau^*_{\bar{Q}}(Y),\zeta) \rangle=\langle \tau_{T^{*}\bar{Q}}(Y),T\tau^*_{\bar{Q}}(Y) \rangle+\langle \mu,\zeta \rangle=\langle \theta_{\bar{Q}}(y),Y  \rangle+\langle \mu,\zeta \rangle ,
\end{split}
\end{align}
for any $(q,y,\mu) \in Q\times_{\bar{Q}}\times T^*\bar{Q} \times \G{g}^* \simeq T^*Q$, and any $(q,Y,\mu,\nu,\zeta) \in Q\times_{\bar{Q}}\times TT^*\bar{Q} \times T\G{g}^* \times \G{g}\simeq TT^*Q$, see also \cite[Prop. 3]{yoshimura2009dirac}. Accordingly, the trivialized canonical 1-form may be given by
\begin{equation}\label{hat-theta-Q}
\widehat{\theta_{Q}}:Q\times_{\bar{Q}} T^*\bar{Q}\times \G{g}^*\to Q\times_{\bar{Q}} T^*T^*\bar{Q}\times  T^*\G{g}^*\times \G{g}^*,\qquad (q,y,\mu)\mapsto (q,\theta_{\bar{Q}}(y),\mu,0,\mu),
\end{equation}
which, thus, reduces at once to
\[
\overline{\theta_Q}:T^*\bar{Q}\times_{\bar{Q}} \tilde{\G{g}}^*\to T^*T^*\bar{Q}\times_{\bar{Q}} \widetilde{\G{L}}, \qquad (y,[q,\mu])\mapsto (\theta_{\bar{Q}}(y),[q,\mu,0,\mu]).
\]

\subsubsection*{\rm \bf Trivialization and reduction  of the canonical symplectic 2-form}
We continue with the trivialization and the reduction of the symplectic 2-form $\Om_Q=-d\t_Q \in \Lambda^2(T^*Q)$. Along the lines of \cite[Prop. 3]{yoshimura2009dirac}, we have
\begin{equation}\label{trv-Omega-Q}
\widehat{\Omega_Q}((q,Y_1,\mu,\nu_1,\zeta_1),(q,Y_2,\mu,\nu_2,\zeta_2))=\Omega_{\bar{Q}}(Y_1,Y_2)-((\tau^\ast_{\bar{Q}})^\ast\bar{B}_\mu)(Y_1,Y_2)+\langle\nu_2,\zeta_1\rangle-\langle \nu_1,\zeta_2 \rangle +\langle \mu,[\zeta_2,\zeta_1] \rangle
\end{equation}
for any $(q,Y_1,\mu,\nu_1,\zeta_1),(q,Y_2,\mu,\nu_2,\zeta_2) \in Q\times_{\bar{Q}}\times TT^*\bar{Q} \times T\G{g}^* \times \G{g}\simeq TT^*Q$ with $\tau_{T^*Q}(Y_1)=\tau_{T^*Q}(Y_2)$, where
\begin{equation}\label{B-mu}
B_{\mu}(X_1,X_2):= \left\langle \mu,B(X_1,X_2) \right\rangle
\end{equation}
for any $X_1,X_2 \in \G{X}(Q)$, and $\bar{B}(v_{\pi(q)},w_{\pi(q)}) := B(v_q^h,w_q^h)$ for any $v,w\in T\bar{Q}$. As a result, the reduced canonical 2-form takes the form of
\begin{align}\label{trv-Omega-Q-reduced}
\begin{split}
& \overline{\Omega_Q}((Y_1,[q,\mu,\nu_1,\zeta_1]),(Y_2,[q,\mu,\nu_2,\zeta_2])) = \\
&\hspace{0.5 cm}\Omega_{\bar{Q}}(Y_1,Y_2)-(\tau^*_{\bar{Q}})^*\bar{B}_{[q,\mu]})(Y_1,Y_2)+\left\langle [q,\nu_2],[q,\zeta_1]  \right\rangle-\left\langle [q,\nu_1],[q,\zeta_2] \right\rangle +\left\langle [q,\mu],[q,[\zeta_2,\zeta_1] ]\right\rangle
\end{split}
\end{align}
for any $(Y_1,[q,\mu,\nu_1,\zeta_1]),(Y_2,[q,\mu,\nu_2,\zeta_2]) \in TT^*\bar{Q}\times_{\bar{Q}} \widetilde{\G{K}}$.
\subsection{Trivializations and reductions of the symplectomorphisms}\label{subsect-symplecto}~

Along the way to the trivialization, and then to the reduction, of the Tulczyjew's triplet \eqref{TT}, what remains are the symplectomorphisms of the top line. To be more precise, presenting the trivialization and the reduction of \eqref{alpha} and \eqref{beta} in the present subsection we shall accomplish our main task.

\subsubsection*{\rm \bf Trivialization and Reduction of $\alpha_{Q}$}

It follows at once from the trivialization \eqref{pairing-trv-symbol-126} of the pairing between $TT^*Q$ and $TTQ$, the trivialization \eqref{kappa-Q-trv} of the canonical involution, and the very definition \eqref{alpha} of the symplectomorphism $\a_Q:TT^*Q\to T^*TQ$ that given any $(q,Y,\mu,\nu,\xi) \in Q\times_{\bar{Q}} TT^*\bar{Q} \times T\G{g}^*\times \G{g} \simeq TT^*Q$, and any $(q,U,\xi,\eta,\zeta) \in Q\times_{\bar{Q}} TT\bar{Q} \times T\G{g}\times \G{g} \simeq TTQ$,
\begin{align}\label{cal-widehat-alpha-Q}
\begin{split}
& \langle \widehat{\alpha_{Q}}(q,Y,\mu,\nu,\xi),(q,U,\xi,\eta,\zeta)  \rangle = -\langle (q,Y,\mu,\nu,\xi),\widehat{\kappa_{Q}}(q,U,\xi,\eta,\zeta) \rangle^{\widehat{}} = \\
& -\langle (q,Y,\mu,\nu,\xi),(q,\kappa_{\bar{Q}}(U),\zeta,\eta+\bar{B}(\tau_{T\bar{Q}}(U),T\tau_{\bar{Q}}(U))+[\xi,\zeta],\xi) \rangle^{\widehat{}} = \\
& - \langle Y,\kappa_{\bar{Q}}(U)\rangle^{\widetilde{}} - \langle \mu,\eta+\bar{B}(\tau_{T\bar{Q}}(U),T\tau_{\bar{Q}}(U))+[\xi,\zeta] \rangle - \langle \nu, \zeta\rangle =\\
& \langle \a_{\bar{Q}}(Y)-i_T\bar{B}_\mu,U\rangle - \langle \mu,\eta\rangle + \langle \ad^*_\xi\mu-\nu,\zeta\rangle.
\end{split}
\end{align}
Accordingly, taking into account the compatibility of the pairing \eqref{pairing-tilde}, we may write
\begin{align}\label{hat-alpha}
\begin{split}
& \widehat{\alpha_{Q}}: Q\times_{\bar{Q}} TT^*\bar{Q}\times T\G{g}^* \times \G{g}\to Q\times_{\bar{Q}} T^*T\bar{Q}\times (T\G{g})^*\times \G{g},\\
& \hspace{2cm} (q,Y,\mu,\nu,\xi)\mapsto (q,\alpha_{\bar{Q}}(Y)-i_T\bar{B}_\mu,\ad_\xi^* \mu-\nu,-\mu,\xi). 
\end{split}
\end{align} 
As for the reduction, the similar computations yield
\begin{align}\label{hat-alpha-}
\begin{split}
& \overline{\alpha_{Q}}: TT^*\bar{Q}\times_{\bar{Q}} \widetilde{\G{K}}\to T^*T\bar{Q}\times_{\bar{Q}} \widetilde{\G{K}},  (Y,[q,\mu,\nu,\xi])\mapsto (\alpha_{\bar{Q}}(Y)-i_T\overline{B}_\mu,[q,\ad_{\zeta}^* \mu-\nu,-\mu,\xi]), 
\end{split}
\end{align} 
where $\overline{B}(\tau_{T\bar{Q}}(U),T\tau_{\bar{Q}}(U)) := [q,\bar{B}(\tau_{T\bar{Q}}(U),T\tau_{\bar{Q}}(U))] \in \tilde{\G{g}}$.

\subsubsection*{\rm \bf Trivialization and Reduction of $\Omega_Q^\flat$}

Recalling the trivialization \eqref{trv-Omega-Q} of the symplectic 2-form, given any $(q,Y_1,\mu,\nu_1,\zeta_1),(q,Y_2,\mu,\nu_2,\zeta_2)  \in Q\times_{\bar{Q}}TT^*\bar{Q}\times (T\G{g})^*\times \G{g} \simeq TT^*Q$, we have
\begin{align}\label{calc-trv-Omega-Q-flat}
\begin{split}
& \langle \widehat{\Omega_Q}^\flat(q,Y_1,\mu,\nu_1,\zeta_1),(q,Y_2,\mu,\nu_2,\zeta_2) \rangle =  \widehat{\Omega_Q}( (q,Y_1,\mu,\nu_1,\zeta_1),(q,Y_2,\mu,\nu_2,\zeta_2) ) = \\
& \Omega_{\bar{Q}}(Y_1,Y_2)-((\tau^*_{\bar{Q}})^*\bar{B}_\mu)(Y_1,Y_2)+\langle \nu_2,\zeta_1  \rangle-\langle \nu_1,\zeta_2 \rangle +\langle \mu,[\zeta_2,\zeta_1] \rangle=\\
& \langle\Omega_{\bar{Q}}^\flat(Y_1)-((\tau^*_{\bar{Q}})^*\bar{B}_\mu)^\flat(Y_1),Y_2\rangle+\langle \nu_2,\zeta_1  \rangle +\langle  \ad^*_{\zeta_1}\mu-\nu_1,\zeta_2\rangle.
\end{split}
\end{align}
In other words, the symplectomorphism $\Omega_Q^\flat:TT^*Q\to T^*T^*Q$ trivializes into
\begin{align}\label{widehat-Omega-Q-flat}
\begin{split}
&\widehat{\Omega_Q}^\flat : Q\times_{\bar{Q}} TT^*\bar{Q}\times T\G{g}^* \times \G{g} \to Q\times_{\bar{Q}} T^*T^*\bar{Q}\times T^*\G{g}^* \times \G{g}^*,\\
&(q,Y,\mu,\nu,\zeta) \mapsto (q,\Omega_{\bar{Q}}^\flat(Y)- ((\tau^*_{\bar{Q}})^*\bar{B}_\mu)^\flat(Y),\mu, \zeta, \ad^*_{\zeta}\mu - \nu  ).
\end{split}
\end{align}
As a result, the $G$-action reduces \eqref{widehat-Omega-Q-flat} into
\begin{align}\label{widehat-Omega-Q-flat-reduced}
\begin{split}
&\overline{\Omega_Q}^\flat :  TT^*\bar{Q}\times_{\bar{Q}} \widetilde{\G{K}} \to T^*T^*\bar{Q}\times_{\bar{Q}} \widetilde{\G{L}},\\
&(Y,[q,\mu,\nu,\zeta]) \mapsto (\Omega_{\bar{Q}}^\flat(Y)- ((\tau^*_{\bar{Q}})^*\bar{B}_\mu)^\flat(Y),[q,\mu, \zeta, \ad^*_{\zeta}\mu - \nu]).
\end{split}
\end{align}

\subsection{Trivialization of Tulczyjew's symplectic space}\label{subsect-symplect-sp}~

In this subsection we shall present the trivialization of the symplectic 2-form on $TT^*Q$. To this end, we shall first study the tangent space $TTT^*Q$ of Tulczyjew's symplectic space from the point of view of the trivialization. Taking into account, then, the operations \eqref{i_T} and \eqref{d-T}, we shall succeed in obtaining the explicit expression of the trivialized symplectic structure on $TT^*Q$. As for the reduction, we shall be content with a brief comment at the end of the section.

\subsubsection*{\rm \bf Trivialization and reduction of $TTT^*Q$}

In view of the trivialization $TT^*Q \simeq Q\times_{\bar{Q}} TT^*\bar{Q}\times T\G{g}^* \times \G{g}$, we have at once
\[
TTT^*Q \simeq TQ\times_{T\bar{Q}} TTT^*\bar{Q}\times TT\G{g}^* \times T\G{g},
\]
which, by $TQ\simeq Q\times_{\bar{Q}} T\bar{Q} \times \G{g}$ and \eqref{Lambda-TT*}, yields
\begin{align*}
& \lambda_{TTT^*}:TTT^*Q \to Q\times_{\bar{Q}} TTT^*\bar{Q}\times TT\G{g}^* \times T\G{g} \times \G{g}, \\
& \dot{Z} \mapsto ((\tau_Q\circ T\tau_Q\circ TT\tau^*_Q)(\dot{Z}), TTh^*(\dot{Z}), TT{\rm\bf J}_Q(\dot{Z}),TA(TT\tau^*_Q(\dot{Z})),A((T\tau_Q\circ TT\tau^*_Q)(\dot{Z})), \\
& \lambda_{TTT^*}^{-1}:Q\times_{\bar{Q}} TTT^*\bar{Q}\times TT\G{g}^* \times T\G{g} \times \G{g} \to TTT^*Q, \\
& (q,\acute{Y},\mu,\nu,\dot{\mu},\dot{\nu},\zeta,\dot{\zeta},\d) \mapsto T_{TT^*\pi((\tau_{T^*\bar{Q}}\circ \tau_{TT^*\bar{Q}})(\acute{Y}))+ {\rm \bf J}^*_{T^*Q}(\d)(q)}TT^*\pi(\acute{Y}) + TTA^*(\mu,\nu,\dot{\mu},\dot{\nu})  + T{\rm \bf J}^*_{T^*Q}(\zeta,\dot{\zeta}) ,
\end{align*}
for any $\dot{Z} \in TTT^*Q$, and any $(q,\acute{Y},\mu,\nu,\dot{\mu},\dot{\nu},\zeta,\dot{\zeta},\d) \in Q\times_{\bar{Q}} TTT^*\bar{Q}\times TT\G{g}^* \times T\G{g} \times \G{g}$. 

In this case, the projections $\tau_{TT^*Q}:TTT^*Q\to TT^*Q$ and $T\tau_{T^*Q}:TTT^*Q\to TT^*Q$ take the form 
\begin{align}\label{proj-triv-TTT*Q}
\begin{split}
& \widehat{\tau_{TT^*Q}}: Q\times_{\bar{Q}} TTT^*\bar{Q}\times TT\G{g}^* \times T\G{g} \times \G{g} \to Q\times_{\bar{Q}} TT^*\bar{Q}\times T\G{g}^* \times \G{g} , \\
&  (q,\acute{Y},\mu,\nu,\dot{\mu},\dot{\nu},\zeta,\dot{\zeta},\d) \mapsto (q,\tau_{TT^*\bar{Q}}(\acute{Y}),\mu,\nu,\zeta), \\
& \widehat{T\tau_{T^*Q}}:Q\times_{\bar{Q}} TTT^*\bar{Q}\times TT\G{g}^* \times T\G{g} \times \G{g} \to Q\times_{\bar{Q}} TT^*\bar{Q}\times T\G{g}^* \times \G{g} , \\
& (q,\acute{Y},\mu,\nu,\dot{\mu},\dot{\nu},\zeta,\dot{\zeta},\d) \mapsto (q,T\tau_{T^*\bar{Q}}(\acute{Y}),\mu,\dot{\mu},\d),
\end{split}
\end{align}
respectively.

Let us next consider the $G$-action
\[
G\times TTT^*Q\to TTT^*Q, \qquad g\cdot \dot{Z}:=TTT^*\phi_{g^{-1}}(\dot{Z}),
\]
which, on the level of the trivialization, appears as
\begin{align*}
& G\times (Q\times_{\bar{Q}} TTT^*\bar{Q}\times TT\G{g}^* \times T\G{g} \times \G{g} ) \to Q\times_{\bar{Q}} TTT^*\bar{Q}\times TT\G{g}^* \times T\G{g} \times \G{g} , \\
& (g,(q,\acute{Y},\mu,\nu,\dot{\mu},\dot{\nu},\zeta,\dot{\zeta},\d)) \mapsto (g\cdot q,\acute{Y},\Ad^*_g\mu,\Ad^*_g\nu,\Ad^*_g\dot{\mu},\Ad^*_g\dot{\nu},\Ad_g\zeta,\Ad_g\dot{\zeta},\Ad_g\d).
\end{align*}
Therefore, 
\[
G\backslash TTT^*Q \simeq TTT^*\bar{Q} \times_{\bar{Q}} \widetilde{\G{M}},
\]
where $\widetilde{\G{M}} := G\backslash(Q\times \G{g}^*\times \G{g}^*\times\G{g}^*\times\G{g}^*\times\G{g}\times\G{g}\times\G{g} )\simeq \widetilde{\G{g}}^*\times_{\bar{Q}} \widetilde{\G{g}}^*\times_{\bar{Q}}\widetilde{\G{g}}^*\times_{\bar{Q}}\widetilde{\G{g}}^*\times_{\bar{Q}}\widetilde{\G{g}}\times_{\bar{Q}}\widetilde{\G{g}}\times_{\bar{Q}}\widetilde{\G{g}}$ through
\begin{align*}
& G\times (Q\times \G{g}^*\times \G{g}^*\times\G{g}^*\times\G{g}^*\times\G{g}\times\G{g}\times\G{g}) \to Q\times \G{g}^*\times \G{g}^*\times\G{g}^*\times\G{g}^*\times\G{g}\times\G{g}\times\G{g}, \\
& (g,(q,\mu,\nu,\dot{\mu},\dot{\nu},\zeta,\dot{\zeta},\d)) \mapsto (g\cdot q,\Ad^*_g\mu,\Ad^*_g\nu,\Ad^*_g\dot{\mu},\Ad^*_g\dot{\nu},\Ad_g\zeta,\Ad_g\dot{\zeta},\Ad_g\d).
\end{align*}
As a result, \eqref{proj-triv-TTT*Q} reduces to
\begin{align}\label{reduction-TTT^*Q-to-TT^*Q}
\begin{split}
& \overline{\tau_{TT^*Q}}: TTT^*\bar{Q}\times_{\bar{Q}}  \widetilde{\G{M}}\to TT^*\bar{Q}\times_{\bar{Q}}  \widetilde{\G{K}} ,\\  &(\acute{Y},[q,\mu,\nu,\dot{\mu},\dot{\nu},\zeta,\dot{\zeta},\d]) \mapsto (\tau_{TT^*\bar{Q}}(\acute{Y}),[q,\mu,\nu,\zeta]), \\
& \overline{T\tau_{T^*Q}}:TTT^*\bar{Q}\times_{\bar{Q}}  \widetilde{\G{M}}\to TT^*\bar{Q}\times_{\bar{Q}}  \widetilde{\G{K}},\\ &(\acute{Y},[q,\mu,\nu,\dot{\mu},\dot{\nu},\zeta,\dot{\zeta},\d]) \mapsto (T\tau_{T^*\bar{Q}}(\acute{Y}),[q,\mu,\dot{\mu},\d]).
\end{split}
\end{align}

\subsubsection*{\rm \bf Trivialization of the symplectic 2-form on Tulczyjew's symplectic space}

We shall, finally, present the trivialization $\widehat{d_T}\widehat{\Om_Q}\in \Lambda^2(Q\times_{\bar{Q}} TT^*\bar{Q}\times T\G{g}^*\times \G{g})$ of the symplectic 2-form $d_T\Om_Q\in\Lambda^2(TT^*Q)$, where the trivialization 
\begin{equation}\label{hat-d_T}
\widehat{d_T}:\Lambda^k(Q)\to \Lambda^k(Q\times_{\bar{Q}} T\bar{Q}\times \G{g}),\qquad \widehat{d_T}=d\widehat{i_T}+\widehat{i_T}d
\end{equation}	
of the operator \eqref{d-T} is given by the trivialization
\begin{equation}\label{widehat-i-T}
\begin{split}
&\widehat{i_T} :\Lambda^k(Q) \to \Lambda^{k-1}(Q\times_{\bar{Q}} T\bar{Q}\times\G{g}),\hspace{0.5 cm} \widehat{i_T}\Omega(X_1,\cdots,X_{k-1})=\Omega(\widehat{\tau_{TQ}}(X_1),\widehat{T\tau_{Q}}(X_1),\cdots,\widehat{T\tau_{Q}}(X_{k-1}) )
\end{split}
\end{equation}
of \eqref{i_T}, for any $X_1,\ldots,X_{k-1} \in \G{X}(Q\times_{\bar{Q}} T\bar{Q}\times\G{g})$ and any $\Om\in \Lambda^k(Q)$.

Accordingly, given any $(q,\acute{Y}_1,\mu,\nu,\dot{\mu}_1,\dot{\nu}_1,\zeta,\dot{\zeta}_1,\d_1),(q,\acute{Y}_2,\mu,\nu,\dot{\mu}_2,\dot{\nu}_2,\zeta,\dot{\zeta}_2,\d_2) \in Q\times_{\bar{Q}} TTT^*\bar{Q}\times TT\G{g}^* \times T\G{g} \times \G{g}$, with $\tau_{TT^*\bar{Q}}(\acute{Y}_1) = \tau_{TT^*\bar{Q}}(\acute{Y}_2) =:Y \in TT^*\bar{Q}$, we have
\begin{align}\label{triv-Om-I}
\begin{split}
& \widehat{d_T}\widehat{\Om_Q} ((q,\acute{Y}_1,\mu,\nu,\dot{\mu}_1,\dot{\nu}_1,\zeta,\dot{\zeta}_1,\d_1),(q,\acute{Y}_2,\mu,\nu,\dot{\mu}_2,\dot{\nu}_2,\zeta,\dot{\zeta}_2,\d_2) ) = \\
& d\widehat{\vartheta}_1 ((q,\acute{Y}_1,\mu,\nu,\dot{\mu}_1,\dot{\nu}_1,\zeta,\dot{\zeta}_1,\d_1),(q,\acute{Y}_2,\mu,\nu,\dot{\mu}_2,\dot{\nu}_2,\zeta,\dot{\zeta}_2,\d_2) ) = \\
&  d\widehat{\vartheta}_1 (X_1,X_2) (q,Y,\mu,\nu,\zeta) = \Big(X_1(\widehat{\vartheta}_1(X_2)) - X_2(\widehat{\vartheta}_1(X_1)) - \widehat{\vartheta}_1([X_1,X_2])\Big)(q,Y,\mu,\nu,\zeta),
\end{split}
\end{align}
where on the first equality we used \eqref{hat-d_T}, $\widehat{\Om_Q}=-d\widehat{\t_Q} \in \Lambda^2(Q\times_{\bar{Q}}T^*\bar{Q}\times \G{g}^*)$, and $\widehat{\vartheta}_1=-\widehat{i_T}\widehat{\Om_Q}$, while  $X_1,X_2\in \G{X}(Q\times_{\bar{Q}}TT^*\bar{Q}\times T\G{g}^*\times \G{g})$ are vector fields given by
\begin{align*}
& X_1(q,Y,\mu,\nu,\zeta) = (q,\acute{Y}_1,\mu,\nu,\dot{\mu}_1,\dot{\nu}_1,\zeta,\dot{\zeta}_1,\d_1) \in Q\times_{\bar{Q}}TTT^*\bar{Q}\times TT\G{g}^*\times T\G{g}\times \G{g}, \\
& X_2(q,Y,\mu,\nu,\zeta) = (q,\acute{Y}_2,\mu,\nu,\dot{\mu}_2,\dot{\nu}_2,\zeta,\dot{\zeta}_2,\d_2) \in Q\times_{\bar{Q}}TTT^*\bar{Q}\times TT\G{g}^*\times T\G{g}\times \G{g}.
\end{align*} 
Fixing $(q,Y,\mu,\nu,\zeta) \in Q\times_{\bar{Q}}TT^*\bar{Q}\times T\G{g}^*\times \G{g}$, let $\phi_1(t):=(q_1(t),Y_1(t),\mu_1(t),\nu_1(t),\zeta_1(t)) \in Q\times_{\bar{Q}}TT^*\bar{Q}\times T\G{g}^*\times \G{g}$ be the integral curve of $X_1 \in \G{X}(Q\times_{\bar{Q}}TT^*\bar{Q}\times T\G{g}^*\times \G{g})$, and $\phi_2(t):=(q_2(t),Y_2(t),\mu_2(t),\nu_2(t),\zeta_2(t)) \in Q\times_{\bar{Q}}TT^*\bar{Q}\times T\G{g}^*\times \G{g}$ of $X_2 \in \G{X}(Q\times_{\bar{Q}}TT^*\bar{Q}\times T\G{g}^*\times \G{g})$, that is,
\begin{align*}
& q_1(0)=q=q_2(0), \qquad \dot{q}_1(0)=\d_1, \qquad \dot{q}_2(0)=\d_2, \\
& \mu_1(0) = \mu = \mu_2(0), \qquad \dot{\mu}_1(0)=\dot{\mu}_1, \qquad \dot{\mu}_2(0)=\dot{\mu}_2, \\
& \zeta_1(0) = \zeta = \zeta_2(0), \qquad \dot{\zeta}_1(0)=\dot{\zeta}_1, \qquad \dot{\zeta}_2(0)=\dot{\zeta}_2, \\
& Y_1(0) = Y = Y_2(0), \qquad \dot{Y}_1(0) = \acute{Y}_1, \qquad \dot{Y}_2(0) = \acute{Y}_2.
\end{align*}
We thus have
\[
X_1(\widehat{\vartheta}_1(X_2))(q,Y,\mu,\nu,\zeta) = \dt \widehat{\vartheta}_1(X_2)(\phi_1(t)) = -\dt \widehat{\Om_Q}(\widehat{\tau_{TT^*Q}}(X_2(\phi_1(t))), \widehat{T\tau_{T^*Q}}(X_2(\phi_1(t)))),
\] 
where we shall write
\[
X_2(\phi_1(t)) = (q_1(t),\acute{Y}_{12}(t),\mu_1(t),\nu_1(t),\dot{\mu}_2,\dot{\nu}_2,\zeta_1(t),\dot{\zeta}_2,\d_2)
\]
for some $\acute{Y}_{12}(t) \in TTT^*\bar{Q}$ satisfying $\acute{Y}_{12}(0)=\acute{Y}_{2}$. Accordingly,
\begin{align}\label{triv-Om-II}
\begin{split}
& X_1(\widehat{\vartheta}_1(X_2))(q,Y,\mu,\nu,\zeta) = \\
& -\dt \widehat{\Om_Q}((q_1(t),\tau_{TT^*Q}(\acute{Y}_{12}(t)),\mu_1(t),\nu_1(t),\zeta_1(t)), (q_1(t),T\tau_{T^*Q}(\acute{Y}_{12}(t)),\mu_1(t),\dot{\mu}_2,\d_2)) = \\
&-\dt \Big( \Om_{\bar{Q}}(\tau_{TT^*Q}(\acute{Y}_{12}(t)),T\tau_{T^*Q}(\acute{Y}_{12}(t)) - ((\tau^*_{\bar{Q}})^*\bar{B}_{\mu_1(t)})(\tau_{TT^*Q}(\acute{Y}_{12}(t)),T\tau_{T^*Q}(\acute{Y}_{12}(t))) + \\
& \hspace{2cm} \langle \dot{\mu}_2,\zeta_1(t)\rangle - \langle \nu_1(t),\d_2\rangle + \langle \mu_1(t),[\d_2,\zeta_1(t)]\rangle\Big) = \\
&-\acute{Y}_1(i_T\Om_{\bar{Q}}(\acute{Y}_2))+ \acute{Y}_1(i_T((\tau^*_{\bar{Q}})^*\bar{B}_{\mu})(\acute{Y}_2)) + i_T((\tau^*_{\bar{Q}})^*\bar{B}_{\dot{\mu}_1})(\acute{Y}_2)-\langle \dot{\mu}_2,\dot{\zeta_1}\rangle + \langle \dot{\nu}_1,\d_2\rangle - \langle \dot{\mu}_1,[\d_2,\zeta_1]\rangle- \langle \mu_1,[\d_2,\dot{\zeta}_1]\rangle .
\end{split}
\end{align}
Similarly, setting
\[
X_1(\phi_2(t)) = (q_2(t),\acute{Y}_{21}(t),\mu_2(t),\nu_2(t),\dot{\mu}_1,\dot{\nu}_1,\zeta_2(t),\dot{\zeta}_1,\d_1)
\]
for some $\acute{Y}_{21}(t) \in TTT^*\bar{Q}$ with $\acute{Y}_{21}(0)=\acute{Y}_2$, we arrive at   
\begin{align}\label{triv-Om-III}
\begin{split}
& X_2(\widehat{\vartheta}_1(X_1))(q,Y,\mu,\nu,\zeta) = \\
&-\acute{Y}_2(i_T\Om_{\bar{Q}}(\acute{Y}_1))+ \acute{Y}_2(i_T((\tau^*_{\bar{Q}})^*\bar{B}_{\mu})(\acute{Y}_1))+   i_T((\tau^*_{\bar{Q}})^*\bar{B}_{\dot{\mu}_2})(\acute{Y}_1) -\langle \dot{\mu}_1,\dot{\zeta_2}\rangle + \langle \dot{\nu}_2,\d_1\rangle - \langle \dot{\mu}_2,[\d_1,\zeta_2]\rangle- \langle \mu_2,[\d_1,\dot{\zeta}_2]\rangle .
\end{split}
\end{align}

Finally, in view of the decomposition $TTT^*Q \simeq V(TT^*Q)\oplus H(TT^*Q)$ of $TTT^*Q$ into the vertical subbundle $V(TT^*Q)=\ker T\tau_{T^*Q} $ and the horizontal subbundle $H(TT^*Q)$, we have, along the lines of the proof of \cite[Prop. 3]{yoshimura2009dirac},
\begin{align}\label{triv-Om-IV}
\begin{split}
& \widehat{\vartheta}_1([X_1,X_2])(q,Y,\mu,\nu,\zeta) = -\widehat{\Om_Q}(\widehat{\tau_{TT^*Q}}[X_1,X_2],\widehat{T\tau_{T^*Q}}[X_1,X_2]) = \\
& -\widehat{\Om_Q}((q,\tau_{TT^*\bar{Q}}[\acute{Y}_1,\acute{Y}_2],\mu,\nu,\zeta),(q,T\tau_{T^*Q}[\acute{Y}_1,\acute{Y}_2],\mu,0,[\d_1,\d_2]-B(T\tau_{T^*\bar{Q}}(\acute{Y}_1),T\tau_{T^*\bar{Q}}(\acute{Y}_2))) = \\
& \hspace{1cm} -\Om_{\bar{Q}}(\tau_{TT^*\bar{Q}}([\acute{Y}_1,\acute{Y}_2]),T\tau_{T^*\bar{Q}}([\acute{Y}_1,\acute{Y}_2])) + (\tau^*)_{\bar{Q}}^*\bar{B}_\mu(\tau_{TT^*\bar{Q}}([\acute{Y}_1,\acute{Y}_2]),T\tau_{T^*\bar{Q}}([\acute{Y}_1,\acute{Y}_2])) + \\
& \langle  \ad^*_{[\d_1,\d_2]-B(T\tau_{T^*\bar{Q}}(\acute{Y}_1),T\tau_{T^*\bar{Q}}(\acute{Y}_2))}\mu, \zeta\rangle + \langle \nu,[\d_1,\d_2]-B(T\tau_{T^*\bar{Q}}(\acute{Y}_1),T\tau_{T^*\bar{Q}}(\acute{Y}_2))\rangle.
\end{split}
\end{align}
Combining \eqref{triv-Om-I}-\eqref{triv-Om-IV} we obtain the trivialization of the symplectic structure on the Tulczyjew's symplectic space as
\begin{align*}
& \widehat{d_T}\widehat{\Om_Q} ((q,\acute{Y}_1,\mu,\nu,\dot{\mu}_1,\dot{\nu}_1,\zeta,\dot{\zeta}_1,\d_1),(q,\acute{Y}_2,\mu,\nu,\dot{\mu}_2,\dot{\nu}_2,\zeta,\dot{\zeta}_2,\d_2) ) = \\
& -d_T\Om_{\bar{Q}}(\acute{Y}_1,\acute{Y}_2)+ di_T((\tau^*_{\bar{Q}})^*\bar{B}_{\dot{\mu}_1})(\acute{Y}_1,\acute{Y}_2) + i_T((\tau^*_{\bar{Q}})^*\bar{B}_{\dot{\mu}_1})(\acute{Y}_2) - i_T((\tau^*_{\bar{Q}})^*\bar{B}_{\dot{\mu}_2})(\acute{Y}_1) +\\
& \langle \ad^*_{\d_2}\mu_1-\dot{\mu}_2,\dot{\zeta_1}\rangle + \langle \dot{\nu}_1-\ad^*_{\zeta_1}\dot{\mu}_1,\d_2\rangle + \langle \dot{\mu}_1-\ad^*_{\d_1}\mu_2,\dot{\zeta_2}\rangle + \langle \ad^*_{\zeta_2}\dot{\mu}_2-\dot{\nu}_2,\d_1\rangle +\\
& \langle  \ad^*_{B(T\tau_{T^*\bar{Q}}(\acute{Y}_1),T\tau_{T^*\bar{Q}}(\acute{Y}_2))-[\d_1,\d_2]}\mu, \zeta\rangle + \langle \nu,B(T\tau_{T^*\bar{Q}}(\acute{Y}_1),T\tau_{T^*\bar{Q}}(\acute{Y}_2))-[\d_1,\d_2])\rangle.
\end{align*}

\begin{remark}
The reduction $\overline{d_T}\overline{\Om_Q}\in\Lambda^2(TT^*\bar{Q} \times_{\bar{Q}} \widetilde{\G{K}})$ of $d_T\Om_Q \in\Lambda^2(TT^*Q)$, under the $G$-action, then follows at once in view of \eqref{reduction-TTT^*Q-to-TT^*Q} with
\[
\overline{d_T}:\Lambda^k(\bar{Q})\to \Lambda^k(T\bar{Q}\times_{\bar{Q}} \tilde{\G{g}}), \qquad \overline{d_T}:= d\overline{i_T} +\overline{i_T} d
\]
and
\[
\overline{i_T} :\Lambda^k(\bar{Q}) \to \Lambda^{k-1}(T\bar{Q}\times_{\bar{Q}} \tilde{\G{g}}),\qquad\overline{i_T}\Omega(X_1,\cdots,X_{k-1})=\Omega(\overline{\tau_{TQ}}(X_1),\overline{T\tau_Q}(X_1),\cdots,\overline{T\tau_Q}(X_{k-1}) )
\]
for any $X_1,\ldots,X_{k-1} \in TT\bar{Q}\times_{\bar{Q}} \tilde{\G{G}}$ and any $\Om\in \Lambda^k(\bar{Q})$, where (by a slight abuse of notation) we mean by $\Lambda^k(\bar{Q})$ the sections of the bundle $\Lambda^k\overline{T^*Q}$ over $\bar{Q}$, and by $\Lambda^{k-1}(T\bar{Q}\times_{\bar{Q}} \tilde{\G{g}})$ the sections of $\Lambda^{k-1}\overline{T^*TQ}$. Accordingly, then, the reduced symplectic form $\overline{d_T}\overline{\Om_Q}\in\Lambda^2(TT^*\bar{Q} \times_{\bar{Q}} \widetilde{\G{K}})$ will be a section of $\Lambda^2\overline{T^*TTQ}$.
\qed
\end{remark}

\section{Conclusion and the Future Work}

In this note, we have presented both the trivialization and the reduction of the Tulczyjew's triplet in the presence of an Ehresmann connection. More precisely, given a manifold admitting a free proper action of a Lie group (which may thus be regarded as a principal bundle over the orbit space of the action), we used the decomposition of its tangent bundle via a principal connection (associated to the principal bundle structure on the manifold through the action) to establish the trivializations of all components of the Tulczyjew's triplet over this manifold. We have then observed, and presented, that once the trivializations are established, it takes rather straightforward calculations to obtain the reductions.

Accordingly, a natural avenue of research is the Legendre transformation of (possibly singular) Lagrange-Poincar\'e systems \cite{cendra2003variational,n2001lagrangian,MaMiOrPeRa07}, which we plan to undertake in a sequel \cite{EsKuSu20}. To this end, we shall need to consider the trivializations and the reductions of Morse families and generating functions, as well as the Lagrangian submanifolds under symmetry. It will then be possible to consider the Legendre transformation even for the singular Lagrangian/Hamiltonian dynamics.

The ultimate goal of our works, \cite{esen2019lifts,esen2011lifts,esen2012geometry,esen2021matched} on the geometry of kinetic theories, and \cite{EsGuSu20,esen2014tulczyjew,esen2015tulczyjew} on the Tulczyjew's triplet for Lie groups, are to establish a Lagrangian formulation of Poisson-Vlasov dynamics of plasma motion. We do hope that the geometry presented herein will help to shed more light on this phenomenon. 

On the other hand, an independent line of research might be to pursue the trivialization and the reduction of Tulczyjew's triplet for the higher order reduced dynamics from the point of view of the higher order Lagrange–Poincaré and Hamilton–Poincaré reductions in \cite{gay2011higher}, see also \cite{gay2012invariant}.

An important application of the present geometry is to study a \textit{gauged} Tulczyjew's triplet motivated by the gauged Lie-Poisson setting of \cite{montgomery1984gauged}, which is constructed on the quotient space $\overline{K\times Q}:=(K\rtimes G)\backslash(K\times Q)$ of the product manifold  $K\times Q$ by the action of the semi-direct product Lie group $K\rtimes G$, where $Q$ is a manifold, whereas $G$ and $K$ are Lie groups. On the level of the dynamical equations, in this case, the additional terms associated to the action of $G$ on $K$ allow to study the dynamics of a particle in a Yang-Mills field \cite{guillemin1980moment,weinstein1978universal}, for nonlinear elasticity \cite{simo1988hamiltonian}, stability of the rigid body, and for Maxwell-Vlasov dynamics of plasma motion \cite{marsden1982hamiltonian}. 
As stated in \cite{montgomery1984gauged}, the gauged geometry is important to establish a passage from the Maxwell-Vlasov  dynamics to Poisson-Vlasov dynamics through the nonrelativistic limit, \cite{Ha09}. For this realization, a motivation behind the need of the semi-direct structure can be found in the work of Van Hove in \cite{van1951probleme}.

\section{Acknowledgment}

The authors would like to thank Prof. Hasan Gümral for his generosity in sharing his expertise on Tulczyjew's triplets. MK acknowledges the excellent research environment in Gebze Technical University, Department of Mathematics, and thanks in particular the department chair Prof. Mansur İsmailov (İsgenderoğlu).  
	MK acknowledges also the support by TUBİTAK (the Scientific and Technological Research Council of Turkey) 2218, National Post-Doctoral Research Fellowship Program.

\bibliographystyle{plain}
\bibliography{references}{}

\def\polhk#1{\setbox0=\hbox{#1}{\ooalign{\hidewidth
  \lower1.5ex\hbox{`}\hidewidth\crcr\unhbox0}}} \def\cprime{$'$}
  \def\cprime{$'$} \def\cprime{$'$} \def\cprime{$'$} \def\cprime{$'$}
  \def\cprime{$'$} \def\cprime{$'$} \def\cprime{$'$} \def\cprime{$'$}
  \def\cprime{$'$} \def\cprime{$'$} \def\Dbar{\leavevmode\lower.6ex\hbox to
  0pt{\hskip-.23ex \accent"16\hss}D}
  \def\cfac#1{\ifmmode\setbox7\hbox{$\accent"5E#1$}\else
  \setbox7\hbox{\accent"5E#1}\penalty 10000\relax\fi\raise 1\ht7
  \hbox{\lower1.15ex\hbox to 1\wd7{\hss\accent"13\hss}}\penalty 10000
  \hskip-1\wd7\penalty 10000\box7}
  \def\cftil#1{\ifmmode\setbox7\hbox{$\accent"5E#1$}\else
  \setbox7\hbox{\accent"5E#1}\penalty 10000\relax\fi\raise 1\ht7
  \hbox{\lower1.15ex\hbox to 1\wd7{\hss\accent"7E\hss}}\penalty 10000
  \hskip-1\wd7\penalty 10000\box7} \def\cprime{$'$}
\begin{thebibliography}{10}

\bibitem{abraham1978foundations}
R.~Abraham and J.~E. Marsden.
\newblock {\em Foundations of mechanics}.
\newblock Benjamin/Cummings Publishing Co., Inc., Advanced Book Program,
  Reading, Mass., 1978.

\bibitem{abrunheiro2018lagrangian}
L.~Abrunheiro and L.~Colombo.
\newblock Lagrangian {L}ie subalgebroids generating dynamics for second-order
  mechanical systems on {L}ie algebroids.
\newblock {\em Mediterr. J. Math.}, 15(2):Paper No. 57, 19, 2018.

\bibitem{arnold1989mathematical}
V.~I. Arnold.
\newblock {\em Mathematical methods of classical mechanics}, volume~60 of {\em
  Graduate Texts in Mathematics}.
\newblock Springer-Verlag, New York, 1989.

\bibitem{barbero2016inverse}
M.~Barbero-Li\~{n}\'{a}n, M.~Farr\'{e}~Puiggal\'{\i}, and D.~Mart\'{\i}n~de
  Diego.
\newblock Inverse problem for {L}agrangian systems on {L}ie algebroids and
  applications to reduction by symmetries.
\newblock {\em Monatsh. Math.}, 180(4):665--691, 2016.

\bibitem{Be11}
S.~Benenti.
\newblock {\em Hamiltonian structures and generating families}.
\newblock Universitext. Springer, New York, 2011.

\bibitem{bloch1996nonholonomic}
A.~M. Bloch, P.~S. Krishnaprasad, J.E. Marsden, and R.~M. Murray.
\newblock Nonholonomic mechanical systems with symmetry.
\newblock {\em Arch. Rational Mech. Anal.}, 136(1):21--99, 1996.

\bibitem{bruce2010tulczyjew}
A.~J. Bruce.
\newblock Tulczyjew triples and higher {P}oisson-{S}chouten structures on {L}ie
  algebroids.
\newblock {\em Rep. Math. Phys.}, 66(2):251--276, 2010.

\bibitem{bruce2015higher}
A.~J. Bruce, K.~Grabowska, and J.~Grabowski.
\newblock Higher order mechanics on graded bundles.
\newblock {\em J. Phys. A}, 48(20):205203, 32, 2015.

\bibitem{cendra2003variational}
H.~Cendra, J.~E. Marsden, S.~Pekarsky, and T.~S. Ratiu.
\newblock Variational principles for {L}ie-{P}oisson and
  {H}amilton-{P}oincar\'{e} equations.
\newblock {\em Mosc. Math. J.}, 3(3):833--867, 1197--1198, 2003.

\bibitem{n2001lagrangian}
H.~Cendra, J.~E. Marsden, and T.~S. Ratiu.
\newblock Lagrangian reduction by stages.
\newblock {\em Mem. Amer. Math. Soc.}, 152(722):x+108, 2001.

\bibitem{de2003tulczyjew}
M.~de~Le{\'o}n, D.~M. de~Diego, and A.~Santamar{\i}a-Merino.
\newblock Tulczyjew’s triples and {L}agrangian submanifolds in classical
  field theory, {A}pplied {D}ifferential {G}eometry and {M}echanics
  ({G}hent)({W}. {S}arlet and {F}. {C}antrijn, {E}ds.), 2003.

\bibitem{deLeLa89}
M.~de~Le\'{o}n and E.~A. Lacomba.
\newblock Lagrangian submanifolds and higher-order mechanical systems.
\newblock {\em J. Phys. A}, 22(18):3809--3820, 1989.

\bibitem{deLeMaMa05}
M.~de~Le\'{o}n, J.~C. Marrero, and E.~Mart\'{\i}nez.
\newblock Lagrangian submanifolds and dynamics on {L}ie algebroids.
\newblock {\em J. Phys. A}, 38(24):R241--R308, 2005.

\bibitem{de2011methods}
M.~de~Le\'{o}n and P.~R. Rodrigues.
\newblock {\em Methods of differential geometry in analytical mechanics},
  volume 158 of {\em North-Holland Mathematics Studies}.
\newblock North-Holland Publishing Co., Amsterdam, 1989.

\bibitem{echeverria2000geometry}
A.~Echeverria-Enr\'{\i}ques, M.~C. Mu\~{n}oz Lecanda, and N.~Rom\'{a}n-Roy.
\newblock Geometry of multisymplectic {H}amiltonian first-order field theories.
\newblock {\em J. Math. Phys.}, 41(11):7402--7444, 2000.

\bibitem{EsGuSu20}
O.~Esen, H.~Gümral, and S.~Sütlü.
\newblock Tulczyjew triplets for {L}ie groups {III} : {H}igher {O}rder
  {D}ynamics and {R}eductions for {I}terated {B}undles.
\newblock (to appear in Theor. Appl. Mech.).

\bibitem{esen2019lifts}
O.~Esen, M.~Grmela, H.~G\"{u}mral, and M.~Pavelka.
\newblock Lifts of symmetric tensors: fluids, plasma, and {G}rad hierarchy.
\newblock {\em Entropy}, 21(9):Paper No. 907, 33, 2019.

\bibitem{esen2011lifts}
O.~Esen and H.~G\"{u}mral.
\newblock Lifts, jets and reduced dynamics.
\newblock {\em Int. J. Geom. Methods Mod. Phys.}, 8(2):331--344, 2011.

\bibitem{esen2012geometry}
O.~Esen and H.~G\"{u}mral.
\newblock Geometry of plasma dynamics {II}: {L}ie algebra of {H}amiltonian
  vector fields.
\newblock {\em J. Geom. Mech.}, 4(3):239--269, 2012.

\bibitem{esen2014tulczyjew}
O.~Esen and H.~G\"{u}mral.
\newblock Tulczyjew's triplet for {L}ie groups {I}: {T}rivializations and
  reductions.
\newblock {\em J. Lie Theory}, 24(4):1115--1160, 2014.

\bibitem{esen2015tulczyjew}
O.~Esen and H.~G\"{u}mral.
\newblock Tulczyjew's triplet for {L}ie groups {II}: {D}ynamics.
\newblock {\em J. Lie Theory}, 27(2):329--356, 2017.

\bibitem{EsKuSu20}
O.~Esen, M.~Kudeyt, and S.~Sütlü.
\newblock Tulczyjew's triplet with an {E}hresmann connection {II}: {D}ynamics.
\newblock (In preparation).

\bibitem{EsenSutl17}
O.~Esen and S.~S\"utl\"u.
\newblock {L}agrangian dynamics on matched pairs.
\newblock {\em J. Geom. Phys.}, 111:142--157, 2017.

\bibitem{esen2021matched}
O.~Esen and S.~S\"{u}tl\"{u}.
\newblock Matched pair analysis of the {V}lasov plasma.
\newblock {\em J. Geom. Mech.}, 13(2):209--246, 2021.

\bibitem{garcia2014reduced}
E.~Garc\'{\i}a-Tora\~{n}o Andr\'{e}s, E.~Guzm\'{a}n, J.~C. Marrero, and
  T.~Mestdag.
\newblock Reduced dynamics and {L}agrangian submanifolds of symplectic
  manifolds.
\newblock {\em J. Phys. A}, 47(22):225203, 24, 2014.

\bibitem{garcia2014geometric}
E.~Garcia Torano~Andres.
\newblock {\em Geometric aspects of reduction for dynamical systems with
  symmetry}.
\newblock PhD thesis, Ghent University, 2014.

\bibitem{gay2012invariant}
F.~Gay-Balmaz, D.~D. Holm, D.~M. Meier, T.~S. Ratiu, and F.-X. Vialard.
\newblock Invariant higher-order variational problems.
\newblock {\em Comm. Math. Phys.}, 309(2):413--458, 2012.

\bibitem{gay2011higher}
F.~Gay-Balmaz, D.~D. Holm, and T.~S. Ratiu.
\newblock Higher order {L}agrange-{P}oincar\'{e} and {H}amilton-{P}oincar\'{e}
  reductions.
\newblock {\em Bull. Braz. Math. Soc. (N.S.)}, 42(4):579--606, 2011.

\bibitem{goldstein1980classical}
H.~Goldstein.
\newblock {\em Classical mechanics}.
\newblock Addison-Wesley Publishing Co., Reading, Mass., second edition, 1980.

\bibitem{grabowska2012tulczyjew}
K.~Grabowska.
\newblock A {T}ulczyjew triple for classical fields.
\newblock {\em J. Phys. A}, 45(14):145207, 35, 2012.

\bibitem{grabowska2013tulczyjew}
K.~Grabowska and J.~Grabowski.
\newblock Tulczyjew triples: from statics to field theory.
\newblock {\em J. Geom. Mech.}, 5(4):445--472, 2013.

\bibitem{grabowska2006geometrical}
K.~Grabowska, P.~Urba\'{n}ski, and J.~Grabowski.
\newblock Geometrical mechanics on algebroids.
\newblock {\em Int. J. Geom. Methods Mod. Phys.}, 3(3):559--575, 2006.

\bibitem{grabowska2015tulczyjew}
K.~Grabowska and L.~Vitagliano.
\newblock Tulczyjew triples in higher derivative field theory.
\newblock {\em J. Geom. Mech.}, 7(1):1--33, 2015.

\bibitem{grabowska2016tulczyjew}
K.~Grabowska and M.~Zaj\c{a}c.
\newblock The {T}ulczyjew triple in mechanics on a {L}ie group.
\newblock {\em J. Geom. Mech.}, 8(4):413--435, 2016.

\bibitem{grabowski2015new}
J.~Grabowski, A.~J. Bruce, K.~Grabowska, and P.~Urba\'{n}ski.
\newblock New developments in geometric mechanics.
\newblock In {\em Geometry of jets and fields}, volume 110 of {\em Banach
  Center Publ.}, pages 57--72. Polish Acad. Sci. Inst. Math., Warsaw, 2016.

\bibitem{GrGrUr14}
J.~Grabowski, K.~Grabowska, and P.~Urba\'{n}ski.
\newblock Geometry of {L}agrangian and {H}amiltonian formalisms in the dynamics
  of strings.
\newblock {\em J. Geom. Mech.}, 6(4):503--526, 2014.

\bibitem{GrabKotoPonc11}
J.~Grabowski, A.~Kotov, and N.~Poncin.
\newblock Geometric structures encoded in the {L}ie structure of an {A}tiyah
  algebroid.
\newblock {\em Transform. Groups}, 16(1):137--160, 2011.

\bibitem{grabowski1997tangent}
J.~Grabowski and P.~Urba\'{n}ski.
\newblock Tangent and cotangent lifts and graded {L}ie algebras associated with
  {L}ie algebroids.
\newblock {\em Ann. Global Anal. Geom.}, 15(5):447--486, 1997.

\bibitem{GrUrRo10}
J.~Grabowski, P.~Urba\'{n}ski, and M.~Rotkiewicz.
\newblock Double affine bundles.
\newblock {\em J. Geom. Phys.}, 60(4):581--598, 2010.

\bibitem{guillemin1980moment}
V.~Guillemin and S.~Sternberg.
\newblock The moment map and collective motion.
\newblock {\em Ann. Physics}, 127(1):220--253, 1980.

\bibitem{Ha09}
H.~G\"{u}mral.
\newblock Geometry of plasma dynamics. {I}. {G}roup of canonical
  diffeomorphisms.
\newblock {\em J. Math. Phys.}, 51(8):083501, 23, 2010.

\bibitem{holm2008geometric}
D.~D. Holm.
\newblock {\em Geometric mechanics. {P}art {I}}.
\newblock Imperial College Press, London, second edition, 2011.
\newblock Dynamics and symmetry.

\bibitem{holm2009geometric}
D.~D. Holm, T.~Schmah, C.~Stoica, and D.~C.~P. Ellis.
\newblock {\em Geometric mechanics and symmetry: from finite to infinite
  dimensions}.
\newblock Oxford University Press London, 2009.

\bibitem{IgMaPaSo06}
D.~Iglesias, J.~C. Marrero, E.~Padr\'{o}n, and D.~Sosa.
\newblock Lagrangian submanifolds and dynamics on {L}ie affgebroids.
\newblock {\em Rep. Math. Phys.}, 57(3):385--436, 2006.

\bibitem{jozwikowski2017prolongations}
M.~J{\'o}{\'z}wikowski.
\newblock Prolongations vs. {T}ulczyjew triples in geometric mechanics.
\newblock {\em arXiv:1712.09858}, 2017.

\bibitem{KobaNomi-book-I}
S.~Kobayashi and K.~Nomizu.
\newblock {\em Foundations of differential geometry. {V}ol {I}}.
\newblock Interscience Publishers, a division of John Wiley \& Sons, New
  York-Lond on, 1963.

\bibitem{KolaMichSlov-book}
I.~Kol{\'a}{\v{r}}, P.~W. Michor, and J.~Slov{\'a}k.
\newblock {\em Natural operations in differential geometry}.
\newblock Springer-Verlag, Berlin, 1993.

\bibitem{LaSnTu75}
B.~Lawruk, J.~\'{S}niatycki, and W.~M. Tulczyjew.
\newblock Special symplectic spaces.
\newblock {\em J. Differential Equations}, 17:477--497, 1975.

\bibitem{libermann2012symplectic}
P.~Libermann and C.-M. Marle.
\newblock {\em Symplectic geometry and analytical mechanics}, volume~35 of {\em
  Mathematics and its Applications}.
\newblock D. Reidel Publishing Co., Dordrecht, 1987.

\bibitem{MaMoRa90}
J.~Marsden, R.~Montgomery, and T.~Ratiu.
\newblock Reduction, symmetry, and phases in mechanics.
\newblock {\em Mem. Amer. Math. Soc.}, 88(436), 1990.

\bibitem{marsden1974reduction}
J.~Marsden and A.~Weinstein.
\newblock Reduction of symplectic manifolds with symmetry.
\newblock {\em Rep. Mathematical Phys.}, 5(1):121--130, 1974.

\bibitem{MaMiOrPeRa07}
J.~E. Marsden, G.~Misio{\l}ek, J.-P. Ortega, M.~Perlmutter, and T.~S. Ratiu.
\newblock {\em {H}amiltonian reduction by stages}, volume 1913 of {\em Lecture
  Notes in Mathematics}.
\newblock Springer, Berlin, 2007.

\bibitem{marsden1986reduction}
J.~E. Marsden and T.~Ratiu.
\newblock Reduction of {P}oisson manifolds.
\newblock {\em Lett. Math. Phys.}, 11(2):161--169, 1986.

\bibitem{MarsdenRatiu-book}
J.~E. Marsden and T.~S. Ratiu.
\newblock {\em Introduction to mechanics and symmetry}, volume~17 of {\em Texts
  in Applied Mathematics}.
\newblock Springer-Verlag, New York, second edition, 1999.

\bibitem{marsden1993lagrangian}
J.~E. Marsden and J.~Scheurle.
\newblock Lagrangian reduction and the double spherical pendulum.
\newblock {\em Z. Angew. Math. Phys.}, 44(1):17--43, 1993.

\bibitem{scheurle1993reduced}
J.~E. Marsden and J.~Scheurle.
\newblock The reduced {E}uler-{L}agrange equations.
\newblock In {\em Dynamics and control of mechanical systems ({W}aterloo, {ON},
  1992)}, volume~1 of {\em Fields Inst. Commun.}, pages 139--164. Amer. Math.
  Soc., Providence, RI, 1993.

\bibitem{marsden1982hamiltonian}
J.~E. Marsden and A.~Weinstein.
\newblock The {H}amiltonian structure of the {M}axwell-{V}lasov equations.
\newblock {\em Phys. D}, 4(3):394--406, 1981/82.

\bibitem{marsden2001comments}
J.~E. Marsden and A.~Weinstein.
\newblock Comments on the history, theory, and applications of symplectic
  reduction.
\newblock In {\em Quantization of singular symplectic quotients}, pages 1--19.
  Springer, 2001.

\bibitem{meyer1973symmetries}
K.~R. Meyer.
\newblock Symmetries and integrals in mechanics.
\newblock In {\em Dynamical systems ({P}roc. {S}ympos., {U}niv. {B}ahia,
  {S}alvador, 1971)}, pages 259--272, 1973.

\bibitem{montgomery1984gauged}
R.~Montgomery, J.~Marsden, and T.~Ratiu.
\newblock Gauged {L}ie-{P}oisson structures.
\newblock In {\em Fluids and plasmas: geometry and dynamics ({B}oulder,
  {C}olo., 1983)}, volume~28 of {\em Contemp. Math.}, pages 101--114. Amer.
  Math. Soc., Providence, RI, 1984.

\bibitem{nakahara2003geometry}
M.~Nakahara.
\newblock {\em Geometry, topology and physics}.
\newblock Graduate Student Series in Physics. Institute of Physics, Bristol,
  second edition, 2003.

\bibitem{Olv-book}
P.~J. Olver.
\newblock {\em Applications of {L}ie groups to differential equations}, volume
  107 of {\em Graduate Texts in Mathematics}.
\newblock Springer-Verlag, New York, second edition, 1993.

\bibitem{olver1995equivalence}
P.~J. Olver.
\newblock {\em Equivalence, invariants and symmetry}.
\newblock Cambridge University Press, 1995.

\bibitem{poincare1901forme}
H.~Poincar{\'e}.
\newblock Sur une forme nouvelle des {\'e}quations de la m{\'e}canique.
\newblock {\em C. R. Acad. Sci. Paris}, 132:369--371, 1901.

\bibitem{roman2007k}
N.~Rom\'{a}n-Roy, \'{A}ngel~M. Rey, M.~Salgado, and S.~Vilari\~{n}o.
\newblock On the {$k$}-symplectic, {$k$}-cosymplectic and multisymplectic
  formalisms of classical field theories.
\newblock {\em J. Geom. Mech.}, 3(1):113--137, 2011.

\bibitem{simo1988hamiltonian}
J.~C. Simo, J.~E. Marsden, and P.~S. Krishnaprasad.
\newblock The {H}amiltonian structure of nonlinear elasticity: the material and
  convective representations of solids, rods, and plates.
\newblock {\em Arch. Rational Mech. Anal.}, 104(2):125--183, 1988.

\bibitem{SnTu72}
J.~\'{S}niatycki and W.~M. Tulczyjew.
\newblock Generating forms of {L}agrangian submanifolds.
\newblock {\em Indiana Univ. Math. J.}, 22:267--275, 1972/73.

\bibitem{tulczyjew1972hamiltonian}
W.~M. Tulczyjew.
\newblock Hamiltonian systems, {L}agrangian systems and the {L}egendre
  transformation.
\newblock In {\em Symposia {M}athematica, {V}ol. {XIV} ({C}onvegno di
  {G}eometria {S}implettica e {F}isica {M}atematica, {INDAM}, {R}ome, 1973)},
  pages 247--258. 1974.

\bibitem{tulczyjew1976soush}
W.~M. Tulczyjew.
\newblock Les sous-vari\'{e}t\'{e}s {L}agrangiennes et la dynamique
  {H}amiltonienne.
\newblock {\em C. R. Acad. Sci. Paris S\'{e}r. A-B}, 283(1):Ai, A15--A18, 1976.

\bibitem{tulczyjew1976sous}
W.~M. Tulczyjew.
\newblock Les sous-vari\'{e}t\'{e}s {L}agrangiennes et la dynamique
  {L}agrangienne.
\newblock {\em C. R. Acad. Sci. Paris S\'{e}r. A-B}, 283(8):Av, A675--A678,
  1976.

\bibitem{Tu77}
W.~M. Tulczyjew.
\newblock The {L}egendre transformation.
\newblock {\em Ann. Inst. H. Poincar\'{e} Sect. A (N.S.)}, 27(1):101--114,
  1977.

\bibitem{Tu80}
W.~M. Tulczyjew.
\newblock A symplectic formulation of relativistic particle dynamics.
\newblock {\em Acta Phys. Polon. B}, 8(6):431--447, 1977.

\bibitem{tulczyjew1989geometric}
W.~M. Tulczyjew.
\newblock {\em Geometric formulations of physical theories}, volume~11 of {\em
  Monographs and Textbooks in Physical Science. Lecture Notes}.
\newblock Bibliopolis, Naples, 1989.
\newblock Statics and dynamics of mechanical systems.

\bibitem{tulczyjew2004homogenous}
W.~M. Tulczyjew and P.~Urba\'{n}ski.
\newblock Homogenous {L}agrangian systems.
\newblock {\em Gravitation, Electromagnetism and Geometric Structures, Pitagora
  Editrice}, 1996.

\bibitem{tulczyjew1999slow}
W.~M. Tulczyjew and P.~Urba\'{n}ski.
\newblock A slow and careful {L}egendre transformation for singular
  {L}agrangians.
\newblock {\em Acta Phys. Polon. B}, 30(10):2909--2978, 1999.

\bibitem{van1951probleme}
L.~Van~Hove.
\newblock Sur le probl\`eme des relations entre les transformations unitaires
  de la m\'{e}canique quantique et les transformations canoniques de la
  m\'{e}canique classique.
\newblock {\em Acad. Roy. Belgique. Bull. Cl. Sci. (5)}, 37:610--620, 1951.

\bibitem{weinstein1978universal}
A.~Weinstein.
\newblock A universal phase space for particles in {Y}ang-{M}ills fields.
\newblock {\em Lett. Math. Phys.}, 2(5):417--420, 1977/78.

\bibitem{We83}
A.~Weinstein.
\newblock The local structure of {P}oisson manifolds.
\newblock {\em J. Differential Geom.}, 18(3):523--557, 1983.

\bibitem{yoshimura2009dirac}
H.~Yoshimura and J.~E. Marsden.
\newblock Dirac cotangent bundle reduction.
\newblock {\em J. Geom. Mech.}, 1(1):87--158, 2009.

\end{thebibliography}

\end{document}